\documentclass[acmsmall]{acmart}

\usepackage{makecell}
\usepackage{bm}
\usepackage{enumitem}
\usepackage{multirow}
\usepackage{float}
\usepackage{tabularx}
\usepackage{pifont}
\usepackage{booktabs}
\usepackage{caption}
\usepackage{colortbl}
\usepackage[subrefformat=parens,skip=0pt]{subcaption}
\usepackage{overpic}
\usepackage{marginnote}
\usepackage[export]{adjustbox}
\usepackage[detect-all]{siunitx}
\usepackage{xspace}

\newcommand{\xmark}{\ding{55}}
\newcommand\ie{i.\,e.\xspace}
\newcommand\eg{e.\,g.\xspace}

\newcommand{\var}[1]{\mathit{#1}}

\def\sym#1{\ifmmode^{#1}\else\(^{#1}\)\fi}

\AtBeginDocument{%
  \providecommand\BibTeX{{%
    \normalfont B\kern-0.5em{\scshape i\kern-0.25em b}\kern-0.8em\TeX}}}

\setcopyright{cc}
\setcctype{by}
\acmJournal{PACMHCI}
\acmYear{2025} \acmVolume{9} \acmNumber{7} \acmArticle{CSCW298} \acmMonth{11} \acmDOI{10.1145/3757479}

\begin{document}

\title{From News Source Sharers to Post Viewers: How Topic Diversity and Conspiracy Theories Shape Engagement With Misinformation During a Health Crisis}

\author{Yuwei Chuai}
\affiliation{
  \institution{University of Luxembourg}
  \country{Luxembourg}
  }
\email{yuwei.chuai@uni.lu}
\orcid{0000-0001-6181-7311}

\author{Jichang Zhao}
\affiliation{
  \institution{Beihang University}
  \country{China}
  }
\email{jichang@buaa.edu.cn}
\orcid{0000-0002-5319-8060}

\author{Gabriele Lenzini}
\affiliation{
  \institution{University of Luxembourg}
  \country{Luxembourg}
  }
\email{gabriele.lenzini@uni.lu}
\orcid{0000-0001-8229-3270}


\begin{abstract}
Online engagement with misinformation threatens societal well-being, particularly during health crises when susceptibility to misinformation is heightened in a multi-topic context. Here, we focus on the COVID-19 pandemic and address a critical gap in understanding engagement with multi-topic misinformation on social media at two user levels: news source sharers (who post news items) and post viewers (who engage with news posts). To this end, we analyze \num{7273} fact-checked source news items and their associated posts on X through the lens of topic diversity and conspiracy theories. We find that false news, especially those containing conspiracy theories, exhibits higher topic diversity than true news. At news source sharer level, false news has a longer lifetime and receives more posts on X than true news, with conspiracy theories further extending its longevity. However, topic diversity does not significantly influence news source sharers' engagement. At post viewer level, contrary to news source sharer level, posts characterized by heightened topic diversity receive more reposts, likes, and replies. Notably, post viewers tend to engage more with misinformation containing conspiracy narratives: false news posts that contain conspiracy theories, on average, receive 40.8\% more reposts, 45.2\% more likes, and 44.1\% more replies compared to those without conspiracy theories. Our findings suggest that news source sharers and post viewers exhibit distinct engagement patterns on X, offering valuable insights into refining misinformation interventions at these two user levels.
\end{abstract}



\received{July 2024}
\received[revised]{December 2024}
\received[accepted]{March 2025}

\maketitle

\section{Introduction}
The spread of misinformation -- \ie, general false or misleading information created either intentionally or unintentionally -- on social media platforms has emerged as one of the most significant global risks in recent years, posing far-reaching consequences for societies and individuals \cite{ecker_psychological_2022,chuai_roll-out_2023}. Health misinformation, compared to misinformation in other domains, carries unique dangers due to its potential to cause harm to human bodies and lives \cite{chen_combating_2022}. This issue is particularly concerning during public health crises, such as the COVID-19 pandemic, when heightened emotions, uncertainty, and increased information consumption create fertile ground for the dissemination of false or misleading information \cite{chuai_anger_2022,solovev_moral_2022,chuai_what_2022,chen_what_2023,freiling_believing_2021}. Additionally, as a distinct type of misinformation, conspiracy theories are deliberately fabricated and attempt to explain the causes of significant public events as secret plots by powerful and malicious groups \cite{douglas_why_2018,kou_conspiracy_2017,suarez-lledo_prevalence_2021,hanley_golden_2023}. Throughout the COVID-19 pandemic, misinformation and conspiracy theories regarding the virus, its origin, and potential treatments or prevention strategies have spread widely on social media, undermining pandemic control efforts \cite{erokhin_covid-19_2022,sharma_covid-19_2022,teng_characterizing_2022}. For instance, widespread anti-vaccine conspiracy theories on social media have strengthened beliefs in the ineffectiveness and side effects of vaccinations, contributing to individuals' vaccination hesitancy and refusal, thereby increasing the risk of COVID-19 infection \cite{sharma_covid-19_2022,pierri_online_2022,enders_relationship_2022}. 

Notably, during health crises, various aspects of society become interconnected. The discussions traverse multiple domains and demonstrate strong interconnectedness, contributing to a rich and diverse informational landscape \cite{miani_interconnectedness_2022,wang_understanding_2022}. For example, a comprehensive analysis of 13.9 million COVID-19 posts on X (formerly Twitter) reveals discussions spanning more than 20 topics across society, encompassing areas such as politics, economy, and racism \cite{chandrasekaran_topics_2020}. Yet, an understanding of how users engage with misinformation within the multi-topic context during health crises is still in its infancy. \citet{vosoughi_spread_2018} examined a large-scale of true and false news posts distributed on X from 2006 to 2017, finding that false news posts are more novel and attract more replies expressing surprise compared to true news posts. The novelty in their study was measured by the topic similarity between the true (false) news post a user reposted and the previous posts to which the user had been exposed before the repost. However, other research shows that social media users often have specific topic preferences and tend to consume similar content, especially within groups of like-minded individuals, \ie, echo chambers \cite{cinelli_echo_2021}. Additionally, social media recommendation algorithms frequently feed users content based on their view histories and increase echo chambers, further reinforcing users' preferences \cite{cinus_effect_2022,engel_learning_2023}. Reconciling these seemingly conflicting findings, one possible explanation is that false news may strategically connect users' interests with other topics to pursue (hidden) agendas. Particularly, in the context of health crises marked by high uncertainty, diverse topics are more interconnected within discussions around crises \cite{chandrasekaran_topics_2020}. Therefore, an analysis of topic interconnectedness in true and false news items during health crises is warranted. 

Previous research has shown that the diversity of topics within information content (\ie, multi-topic information) plays an important role in shaping perceptions and assessments of credibility through the mechanism of associative inference \cite{lee_associative_2023}. Associative inference is an adaptive process that enables individuals to link related information and form novel connections, even in the absence of direct experience, thereby increasing individuals’ susceptibility to online misinformation \cite{lee_associative_2023}. Moreover, the associative inference around online misinformation can evolve within a reinforcing cycle. Individuals collaboratively construct a misleading version of reality and amplify alleged evidence. This facilitates subsequent actions that, in turn, further reinforce the manufactured reality \cite{prochaska_mobilizing_2023}. Therefore, the interconnected and associative multi-topic discussions during health crises underscore the potential role of topic diversity in driving engagement with misinformation on social media. Based on this rationale, we address the following research questions to explore topic diversity in misinformation (RQ1.1) and examine whether it shapes user engagement (RQ1.2):

\begin{itemize}[leftmargin=*]
    \item \textit{RQ1.1: Does misinformation exhibit higher topic diversity compared to true information?}
    \item \textit{RQ1.2: Is topic diversity positively associated with user engagement, and is this association stronger for misinformation than for true information?}
\end{itemize}

Moreover, the integration of conspiracy theories within misinformation narratives serves as an effective strategy to deliberately interconnect various topics \cite{miani_interconnectedness_2022,hanley_golden_2023}. This deliberate approach contributes to an increase in individuals' beliefs in misinformation by fostering illusory pattern perception and promoting associative inference \cite{van_prooijen_connecting_2018,lee_associative_2023}. In the early months of the COVID-19 pandemic, social media platforms were inundated with various conspiracy theories related to the origins, spread, and treatment of the virus \cite{erokhin_covid-19_2022}. The propagation of misinformation with coordinated conspiracies has promoted distrust in public health authorities and skepticism about the safety and effectiveness of COVID-19 vaccines \cite{sharma_covid-19_2022,schmitz_detecting_2023}. While the proliferation of conspiracy theories on social media has raised significant concerns, little is known about how the incorporation of conspiracy theories in misleading narratives differentiates individuals' engagement on social media compared to other types of misinformation \cite{kim_information_2023}. Given this, we further investigate the following research questions to analyze whether misinformation containing conspiracy theories exhibits higher topic diversity (RQ2.1) and receives more engagement (RQ2.2) compared to misinformation without conspiracy theories:

\begin{itemize}[leftmargin=*]
    \item \textit{RQ2.1: Does misinformation containing conspiracy theories exhibit higher topic diversity compared to misinformation without conspiracy theories?}
    \item \textit{RQ2.2: Does misinformation containing conspiracy theories receive more engagement compared to misinformation without conspiracy theories?}
\end{itemize}

To address our research questions, we perform a comprehensive analysis of online engagement with multi-topic misinformation during the early stages of the COVID-19 pandemic, focusing on topic diversity and conspiracy theories. It is worth noting that we examine engagement at two levels: \emph{news source sharer level} and \emph{post viewer level}. News source sharers create posts on X that link to source news items, while post viewers encounter these news posts within the platform and then engage by reposting, liking, or replying. The two user groups may have different motivations for engagement \cite{hanley_golden_2023,teng_characterizing_2022,ballard_conspiracy_2022}. Additionally, while previous studies have extensively explored post viewer engagement on social media \cite{vosoughi_spread_2018,chuai_anger_2022,prollochs_emotions_2021-1,solovev_moral_2022}, research on news source sharer engagement remains limited. Understanding this under-investigated area is crucial to complement our knowledge of news-sharing behavior.

\textbf{Methodology:}
We conduct our analysis based on a COVID-19 dataset consisting of \num{7273} fact-checked source news claims and their corresponding posts on X. These source news claims span six languages and cover the initial period of COVID-19 from January to September 2020. At news source sharer level, the engagement metrics include the number of posts linking to a specific news item and the lifetime of the news item from the creation of its first post to its latest post. At post viewer level, the engagement metrics include repost count, like count, and reply count. To gain a comprehensive understanding of the topic spectrum surrounding COVID-19 discussions, we additionally use over 15 million daily sampled posts at the early stage of COVID-19 to train a Word2Vec model and collect topic keywords. Subsequently, we use the topic keywords to calculate topic diversity and identify conspiracy theories. Finally, we employ multiple regression models to investigate our research questions.

\textbf{Contributions:} 
Our work reveals that elevated topic diversity has been a characteristic of COVID-19 misinformation, particularly for those accompanied by conspiracy narratives. This characteristic becomes more significant within false news posts on X compared to their source news claims. At post viewer level, heightened topic diversity and the integration of conspiracy theories significantly correlate with increased engagement (\ie, reposts, likes, and replies) with misinformation on X. For instance, false news posts that contain conspiracy theories, on average, receive 40.8\% more reposts, 45.2\% more likes, and 44.1\% more replies compared to false news posts without conspiracy theories. However, at news source sharer level, topic diversity and conspiracy theories have no significant associations with the number of posts that source news items receive. This difference suggests that news source sharers and post viewers have distinct engagement patterns. Our findings offer valuable insights into understanding engagement with multi-topic misinformation from news source sharers to post viewers on social media during health crises. Additionally, by highlighting the different engagement patterns at news source sharer and post viewer levels in terms of topic diversity and conspiracy theories, this research informs the development of targeted interventions and strategies at both user levels to mitigate engagement with false information and foster a more resilient society.

\section{Background and Related Work}

\subsection{Misinformation on Social Media}
Social media platforms have transformed information exchange, enabling public users to create and engage with online content globally and instantaneously. However, the absence of quality control mechanisms inherent in traditional media has led to a significant rise in misinformation on social media. Particularly, the spread of misinformation on social media has posed far-reaching consequences across many critical societal domains ranging from politics to public health, thus becoming a concerning global issue in recent years \cite{ecker_psychological_2022,pierri_online_2022,enders_relationship_2022,green_online_2022}. Consequently, social media platforms are under increasing pressure to implement effective measures to reduce the spread of misinformation. Understanding how users engage with misinformation and contribute to its viral spread on social media can help social media platforms design targeted interventions. Prior research has shown that online misinformation spreads farther, faster, deeper, and more broadly compared to true information, as evidenced by the examination of online posts fact-checked by third-party organizations \cite{vosoughi_spread_2018}. Nevertheless, another study, based on community fact-checked posts on X, indicates that misinformation receives fewer reposts compared to true information \cite{drolsbach_diffusion_2023}. This contradiction highlights that not every piece of misinformation spreads faster than true information. Instead, there are likely crucial factors behind the veracity of information that make misinformation unique and contribute to its online diffusion.

Several works have focused on understanding what characterizes misinformation and contributes to its viral spread. For instance, misinformation is more likely to incorporate negative emotions, such as anger, which has been identified as an important driver for information diffusion \cite{chuai_anger_2022,chuai_what_2022,horner_emotions_2021,robertson_negativity_2023}. Additionally, there is a partisan asymmetry that misinformation tends to be right-leaning and promote conservative positions \cite{chuai_political_2023}. Social media users are more likely to believe and share information that is congruent to their political ideologies \cite{garrett_conservatives_2021,rao_partisan_2022,robertson_users_2023,osmundsen_partisan_2021,jenke_affective_2023}. Therefore, right-leaning users are more likely to share misinformation on social media, compared to left-leaning users \cite{nikolov_right_2021}. More importantly, uncertain situations, such as public emergencies or crises, create fertile ground for online misinformation to thrive. During such times, people are eager to know more information about the situation and are thus more likely to believe and share related misinformation, which can undermine trust in control policies and cause detrimental harm to societies and individuals. 

\subsection{Misinformation During Public Events/Crises}

During public events or crises, characterized by heightened emotions, uncertainty, and a surge in information consumption, falsehoods that cater to specific demands and offer related information to contextualize the situation become rampant and easily spread \cite{chuai_anger_2022}. The preservation of democracy is crucial during political elections. However, democracy can be undermined by the widespread dissemination of dedicated online misinformation aimed at pursuing specific political agendas. It has been found that people are exposed to misinformation more often during elections than usual \cite{allcott_social_2017,chuai_political_2023}. Election-related misinformation is often polarized, which can increase people's susceptibility and affect their voting behaviors \cite{rathje_out-group_2021,nikolov_right_2021,recuero_hyperpartisanship_2020}. Additionally, climate change poses challenges at multiple levels, with any mitigation efforts carrying significant economic and political implications \cite{lewandowsky_climate_2021}. Consequently, the debate surrounding global climate change has persisted for years and has become increasingly polarized on social media \cite{fischer_polarized_2022,farrell_growth_2019,falkenberg_growing_2022}. This polarized environment fosters climate change denial and misinformation, further complicating efforts to address and control climate change \cite{lewandowsky_climate_2021, falkenberg_growing_2022}.

The spread of misinformation during health crises is particularly severe, as it can directly harm human health and even cost lives \cite{chen_combating_2022}. Throughout the COVID-19 pandemic, misinformation on a wide variety of COVID-19 related topics has circulated widely on social media platforms. For example, the widespread misinformation on the COVID-19 vaccine reduces users' beliefs in its effectiveness, impeding the fight against the spread of COVID-19 \cite{enders_relationship_2022,loomba_measuring_2021,pierri_online_2022}. A global survey has revealed that a substantial part of people consider COVID-19 misinformation as highly reliable \cite{roozenbeek_susceptibility_2020}. Understanding how misinformation spreads and attracts users during such public events and crises is crucial for developing effective intervention strategies for emergency management.

\subsection{Engagement With COVID-19 Misinformation on Social Media}

The engagement with misinformation during health crises, particularly COVID-19, poses significant threats to crisis control and public safety. Several studies have scrutinized the factors that can influence engagement with COVID-19 misinformation on social media. Recognizing emotions as a significant driver of online information diffusion \cite{chuai_anger_2022,prollochs_emotions_2021,prollochs_emotions_2021-1}, research in this body has predominantly focused on the role of emotions during COVID-19 \cite{solovev_moral_2022,chuai_what_2022,freiling_believing_2021,charquero-ballester_different_2021}. For example, \citet{solovev_moral_2022} conducted a large-scale computational analysis on the role of moral emotions in the spread of COVID-19 misinformation. They found that false posts with higher other-condemning emotions (\eg, anger) are more viral (\ie, more reposts) than true posts. Based on experimental design, \citet{freiling_believing_2021} investigated the role of anxiety in misinformation belief and sharing during COVID-19, revealing that anxiety is a driving factor in believing and sharing both true and false claims. \citet{wang_understanding_2023-1} put the focus on the images and found that online posts containing COVID-19 misinformation images receive similar interactions with posts containing random images. However, COVID-19 misinformation images are shared for longer periods than non-misinformation ones. Additionally, COVID-19 misinformation, mimicking the format and language features of news and scientific reports, is likely to receive more likes \cite{ngai_impact_2022}. For the concern that social bots may share a large amount of misinformation during COVID-19, \citet{teng_characterizing_2022} analyzed different types of users in the engagement with misinformation. They found that the role of social bots in misinformation sharing is limited, and individuals with human-like behavior play a more prominent role.

Knowing the potential role of topics in engagement with misinformation, \citet{ngai_impact_2022} focused on the specific topics in the engagement with COVID-19 vaccine misinformation. They found that safety concerns are the most significant content theme but are negatively associated with likes and shares. However, during COVID-19, various aspects of society became interconnected. COVID-19 misinformation covered a broad spectrum of topics, including public health, politics, medicine, vaccines, and so forth \cite{wang_understanding_2022,chandrasekaran_topics_2020}. The multi-topic misinformation about COVID-19 carries high uncertainty and easily causes confusion to individuals, thus posing threats to people and society \cite{wang_understanding_2022}. It has been revealed that individuals can adaptively link multiple related topics and make novel connections through the mechanism of associative inference, which increases the susceptibility to misinformation \cite{lee_associative_2023}.

\subsection{Multi-Topic Context and Conspiracy Theories During COVID-19}

The discussion on COVID-19 during the pandemic not only focused on the direct health impacts but also spilled over into various social, economic, and political domains. Within the social domain, at the beginning of the pandemic, rumors and conspiracy theories about the origin of the virus in China were rampant on social media, exploiting anti-Asian racism and hate speech on social media \cite{huang_cost_2023}. Using COVID-19 as a weapon, malicious misinformation fostered online hate to spread quickly beyond the control of social media platforms \cite{velasquez_online_2021}. The rising anti-Asian racism during the COVID-19 pandemic knocked Asian businesses \cite{huang_cost_2023,luca_evolution_2024,tang_racial_2023}. In the economic domain, the pandemic hit the global economy, particularly impacting sectors such as travel and leisure \cite{wang_spillover_2023}. Additionally, the scientific uncertainty surrounding the COVID-19 pandemic fueled the weaponization of COVID-19 for politicization \cite{kreps_model_2020}. Some individuals and groups have sought to exploit the pandemic to spread extremist ideologies and radicalize others \cite{davies_witchs_2023}. The high degree of politicization within COVID-19 information contributed to polarization in attitudes towards the pandemic and even to political ideologies \cite{hart_politicization_2020}. For instance, the debate over vaccination became increasingly polarized during the pandemic, expanding far beyond the medical discourse and entering the domain of organizational politics \cite{dehghan_politicization_2022}. Taken together, the COVID-19 pandemic has been incorporated into many other societal domains to pursue specific interests, yielding significant spillover effects on the stability of societies. In this paper, we use topic diversity to denote the extent to which the information involves multiple topics and explore its impact on the engagement with misinformation on social media during COVID-19. 

Particularly, at the onset of the COVID-19 pandemic, a myriad of conspiracy theories emerged, providing alternative explanations for the origins, spread, and treatment of the virus \cite{erokhin_covid-19_2022}. Examples include claims suggesting that the COVID-19 virus is a man-made bioweapon \cite{moffitt_hunting_2021,erokhin_covid-19_2022}. Conspiracy theories often revolve around the notion of secret plots orchestrated by malevolent and powerful groups, deliberately crafted to achieve specific outcomes. This characteristic of conspiracy theories facilitates the connection of multiple topics through an illusory perception pattern, potentially fostering associative inference among individuals  \cite{lee_associative_2023,miani_interconnectedness_2022,van_prooijen_connecting_2018}. The integration of conspiracy theories within misinformation became more pronounced during the COVID-19 pandemic, marked by a significant increase in hyperlinks from misinformation websites to conspiracy theories \cite{hanley_golden_2023}. Despite this observation, the question of whether misinformation containing conspiracy theories receives more engagement than other types of misinformation remains to be explored. In light of these considerations, this paper tries to address the current research gaps by investigating the roles of topic diversity and conspiracy theories in the engagement with misinformation during health crises.

\section{Data and Methods}
\subsection{Dataset: MM-COVID}
The data for our analysis is sourced from a publicly available repository on GitHub \cite{li_mm-covid_2020}. This repository provides an extensive COVID-19 dataset that consists of \num{8825} source news claims and their corresponding engagement on X. Initially, the repository gathers these news claims and their veracity verdicts from popular third-party fact-checking organizations worldwide, including Snopes and Poynter, as well as official health websites. To keep the quantity of the source claims in each language, the repository filters six languages: English, Spanish, Portuguese, Hindi, French, and Italian. This enables a broad and diverse examination of misinformation across linguistic and cultural contexts. Subsequently, online posts that explicitly reference these news sources are collected from X through its advanced search API. Notably, while the news sources in the dataset are linked to the original news URLs, precisely parsing the original news pages and obtaining their content is challenging due to the diverse range of news items collected from various languages and web pages. Instead, the dataset provides news claims, representing the main opinions and topics.

\textbf{Data preprocessing:}
We begin by translating all the non-English texts in source news claims and their associated posts into English using Google Translate API via Googletrans.\footnote{Googletans library is available at \url{https://github.com/ssut/py-googletrans}. It is a free and unlimited Python library that implements Google Translate API and uses the same servers that translate.google.com uses.} Google Translate has a high performance in translating between English and non-English languages, retaining overall meaning in 82.5\% of the translations \cite{taira_pragmatic_2021,sebo_performance_2024}. Subsequently, we perform text preprocessing by removing non-alphabetic and non-numerical characters, stop words, and URLs. We lemmatize all the words using the English transformer pipeline on Spacy and only consider news claims and posts that contain at least one lemmatized word. Additionally, we exclude posts that were posted before January 1, 2020. Due to the absence of release dates for source news items in the dataset, we use the first post that links to the specific source news item as its release date. In the end, we obtain \num{8822} news claims and associated \num{85978} posts. Specifically, \num{59363} posts are linked to \num{1543} false news claims, and \num{26615} posts are linked to \num{7279} true news claims.

\subsection{COVID-19 Topic Spectrum}
A previous study analyzed over 13 million English COVID-19 posts and identified 26 topics within 10 broad themes, including source, prevention, spread and growth, treatment and recovery, impact on the economy and markets, impact on health care sector, and government response, political impact, and racism \cite{chandrasekaran_topics_2020}. This research also provided keywords corresponding to each topic. To ensure a comprehensive understanding of the topic spectrum surrounding COVID-19 discussions, we utilize these keywords as seeds to identify similar words and construct a topic lexicon based on a Word2Vec model. Upon reevaluating the associations between topic categories and their themes, we find that the keywords in the topic ``Hotspots and location'' relate solely to places, such as ``cities'' and ``location,'' not adequately reflecting the theme ``Spread and growth'' in terms of transmission modes or case numbers. Therefore, in the main analysis, we remove the topic ``Hotspots and location'' and focus on the remaining 25 topics across 10 themes (Table \ref{tab:themes_topics}).\footnote{In a subsequent robustness check, we conduct the analysis again using all the original topics and keywords. The results are robust and consistently support our findings. The full estimation results are reported in Suppl. \ref{sec:original_topics}.}

\begin{table}
\centering
\caption{COVID-19 topic spectrum. The 25 topics across 10 broad themes are included.}
\setlength{\tabcolsep}{2pt}
\begin{tabularx}{\columnwidth}{@{\hspace{\tabcolsep}\extracolsep{\fill}}ll}
\toprule
\textbf{Theme} & \textbf{Topic}\\
\midrule
Source (origin)&Outbreak, alternative causes\\
\rowcolor{gray!50}
Prevention&Social distancing, disinfecting and cleanliness\\
Symptoms&Symptoms\\
\rowcolor{gray!50}
Spread and growth&Modes of transmission, spread of cases, death reports\\
Treatment and recovery&Drugs and vaccines, therapies, alternative methods,\\
&testing\\
\rowcolor{gray!50}
Impact on health care sector&Impact on hospitals and clinics, health policy, \\
\rowcolor{gray!50}
&frontline workers\\
Government response&Travel restrictions, financial measures, \\
&lockdown regulations\\
\rowcolor{gray!50}
Impact on the economy and markets&Shortage of products, panic buying, stock markets, \\
\rowcolor{gray!50}
&employment, impact on business\\
Political impacts&Political impacts\\
\rowcolor{gray!50}
Racism&Racism\\
\bottomrule
\end{tabularx}
\label{tab:themes_topics}
\end{table}

\begin{table}
\centering
\caption{The examples of topic seeds (bioweapon, vaccine, election, and quarantine) and their 10 most similar words.}
\begin{tabularx}{\columnwidth}{@{\hspace{\tabcolsep}\extracolsep{\fill}}llll}
\toprule
\multicolumn{1}{l}{\textbf{bioweapon}}&\multicolumn{1}{l}{\textbf{vaccine}}&\multicolumn{1}{l}{\textbf{election}}&\multicolumn{1}{l}{\textbf{quarantine}}\\
\midrule
biowarfare&vaccines&elections&isolation\\
biologicalweapon&vaccination&primaries&quarantined\\
bioweapons&vaccin&voting&selfquarantine\\
bioengineered&vax&2020election&quarentine\\
biologicalwarfare&vacine&electoral&selfisolation\\
bioterrorism&cure&election2020&quarantining\\
engineered&inoculation&referendum&qurantine\\
weaponized&treatments&vote&lockdown\\
wuhanlab&coronavirusvaccine&votes&quarintine\\
biowar&antibodies&generalelection&isolating\\
\bottomrule
\end{tabularx}
\label{tab:word2vec}
\end{table}

\textbf{Word2Vec model:}
To train the Word2Vec model, we additionally collect a large-scale COVID-19 dataset from a public repository on Kaggle.\footnote{The dataset is available at \url{https://www.kaggle.com/datasets/smid80/coronavirus-covid19-tweets}.} This dataset contains \num{26854078} posts in multiple languages, posted from March 9 to April 30, 2020 on X. Given the size of the dataset, we only filter English posts and do not involve the translation process. By combining this dataset with MM-COVID, we obtain a total of \num{15059238} English posts. After removing non-alphabetic and non-numerical characters as well as URLs in the posts, we utilize the refined corpus to train a Word2vec model with a vector size of 200 using the gensim library.\footnote{The model is trained on the High Performance Computing platform at the University of Luxembourg.} The examples of the keywords and their ten most similar words are shown in Table \ref{tab:word2vec}.

\textbf{Topic lexicon:}
We expand our set of topic keywords by incorporating words that exhibit cosine similarities of more than 0.6 with at least one of the initial topic seeds. This process yields 828 additional topic keywords, and they are automatically categorized into the same topics as their corresponding topic seeds. Additionally, we conduct a manual review of these keywords and remove ambiguous terms and abbreviations. For example, we remove the keyword ``reason'' from the topic ``Alternative causes,'' as it does not specify a particular alternative cause for COVID-19, whereas the keyword ``chinavirus'' is an effective term for this topic category. Similarly, we remove ambiguous abbreviations such as ``pres'' in the topic ``Political impacts.'' Ultimately, we retain 90.6\% of the 828 keywords and integrate them with the topic seeds to create an extensive topic lexicon.\footnote{We repeat our analysis using the keywords without manual validation. The results remain consistent, further enhancing the robustness of our findings. The full estimation results are reported in Suppl. \ref{sec:original_topics}.}

\textbf{Topic identification:}
We utilize the topic lexicon to identify topics discussed in both news claims and their associated posts. As a result, 82.4\% of news claims contained at least one topic. Additionally, our analysis considers both the content of the source claims and the content of the corresponding posts when identifying topics in online posts. Consequently, 91.4\% of posts contain at least one topic.

\subsection{Topic Diversity}
\label{sec:topic_diversity}

\begin{figure}[htb]
\centering
\begin{subfigure}{0.32\textwidth}
\centering
\caption{}
\includegraphics[width=\textwidth]{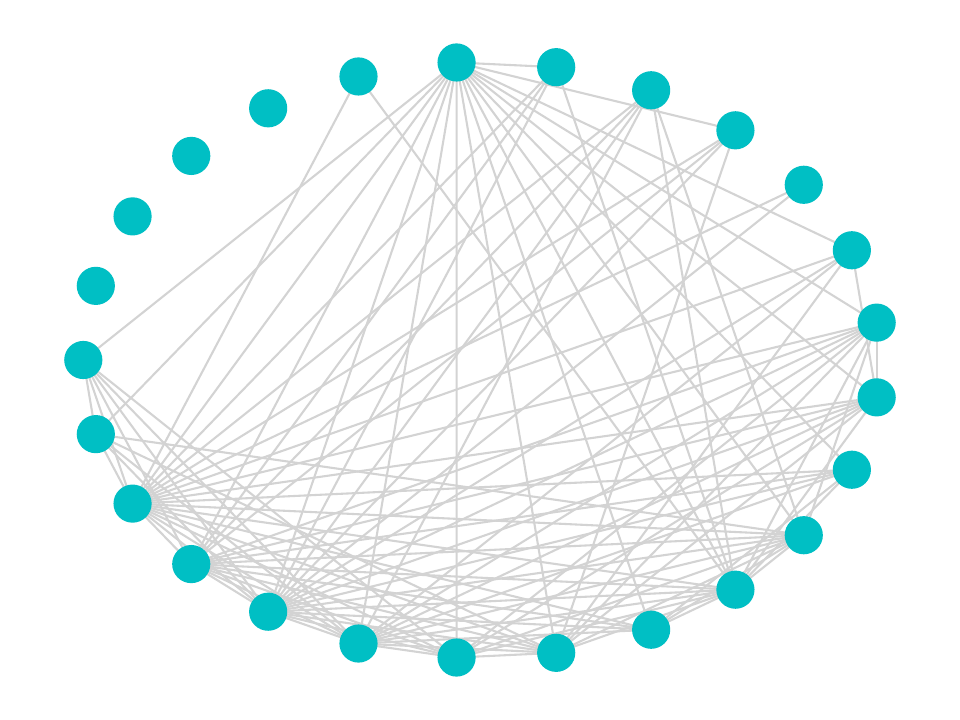}
\label{fig:topic_network_true}
\end{subfigure}
\hspace{10mm}
\begin{subfigure}{0.32\textwidth}
\centering
\caption{}
\includegraphics[width=\textwidth]{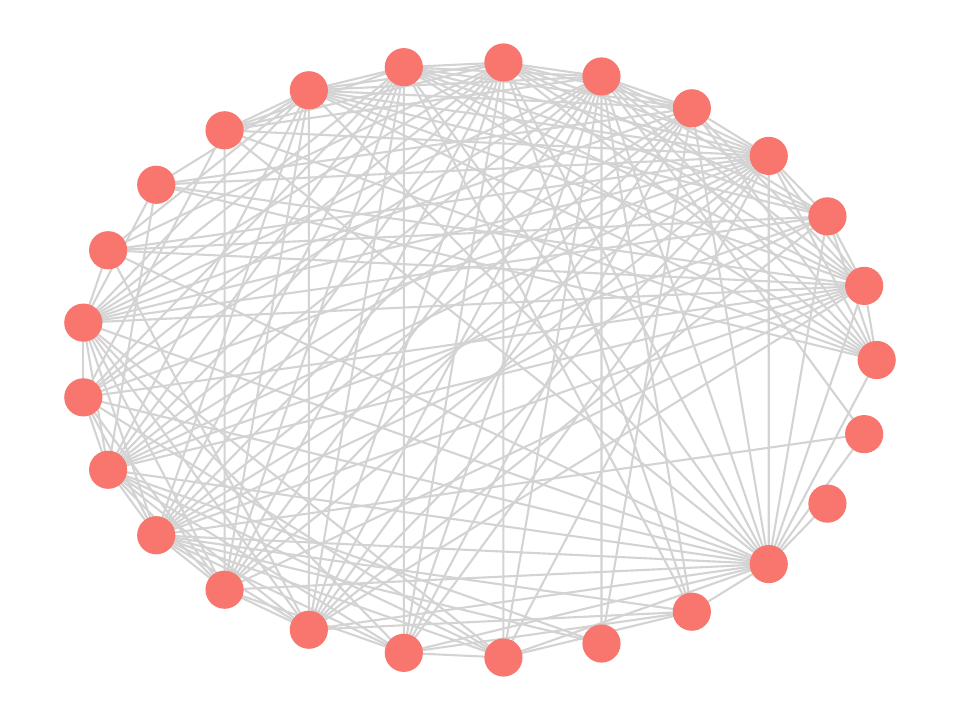}
\label{fig:topic_network_false}
\end{subfigure}

\begin{subfigure}{\textwidth}
\centering
\caption{}
\includegraphics[width=.7\textwidth]{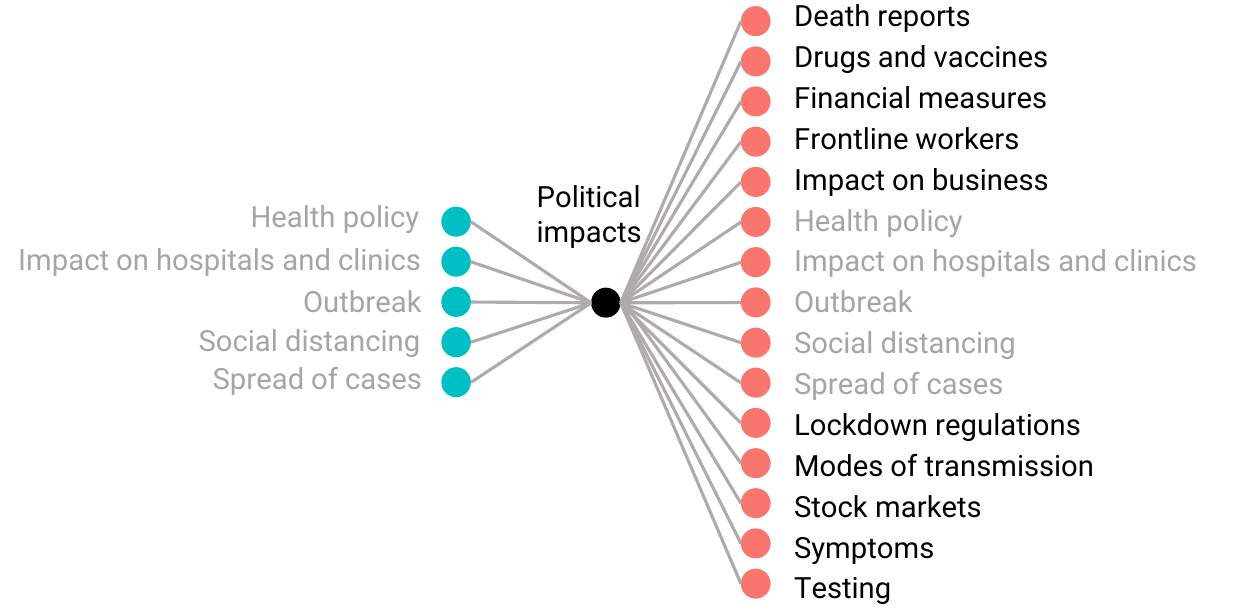}
\label{fig:topic_co_example}
\end{subfigure}
\vspace{1pt}
\caption{Topic co-occurrence in true and false news claims. \subref{fig:topic_network_true} Topic co-occurrence network in true news claims. The number of edges is 107. \subref{fig:topic_network_false} Topic co-occurrence network in false news claims. The number of edges is 167.  \subref{fig:topic_co_example} An example of topic co-occurrences for ``Political impacts.'' Shown are the topic ``Political impacts'' and its co-occurring topics in true and false news claims.}
\label{fig:topic_co}
\end{figure}

Topic diversity indicates the extent to which the content involves multiple topics. To give a sense of topic diversity, we construct topic co-occurrence networks for true and false news claims. Specifically, we calculate the frequency of topic co-occurrence of any two topics out of 25 topics in true and false news claims, and retain the co-occurrence edges with frequencies exceeding the median threshold. In the network of true news claims (Fig. \ref{fig:topic_network_true}), the number of edges is 107, while in the network of false news claims (Fig. \ref{fig:topic_network_false}), the number of edges is 167. Compared to true news claims, false news claims contain 56.1\% more co-occurrence edges and are more interconnected, indicating higher interconnectedness and suggesting greater topic diversity in false news claims. For example, the topic ``Political impacts'' is connected to 5 topics in true news claims, whereas it connects to 10 additional topics, such as ``Death reports'' and ``Drugs and vaccines,'' in false news claims (Fig. \ref{fig:topic_co_example}).

Building upon previous research that used the Gini coefficient to measure the inequality of topic distribution in the content \cite{miani_interconnectedness_2022}, we quantitatively measure topic diversity ($\var{Diversity}$) as:
\begin{equation}
    \var{Diversity} = 1-\var{Gini} = 1 - \frac{\sum_{i=1}^{n} \sum_{j=1}^{n} |x_i - x_j|}{2n^2\bar{x}},
\end{equation}
where $\var{x_{i}}$ denotes the number of words associated to topic $\var{i}$ in each news claim (post), and $\var{n}$ is the number of topics. In our analysis, we exclude news items and posts that do not relate to any topic, resulting in \num{7273} source news claims with corresponding \num{70904} posts (see details in Table \ref{tab:data_summary}).

\subsection{COVID-19 Conspiracy Theories}
The outbreak of COVID-19 has fueled widespread conspiracy theories with misleading claims about its origin, treatment, and other related topics. We summarize the conspiracy theories identified in previous research \cite{erokhin_covid-19_2022,moffitt_hunting_2021,sharma_covid-19_2022} and use a set of keywords and phrases to identify conspiracy theories embedded in false news claims and subsequent posts. Specifically, we consider eight COVID-19 conspiracy theory categories that connect to at least one topic in our analysis: the sources and origins of the virus that caused the outbreak of COVID-19 (VirusOrigin), the connection between 5G and COVID-19 (5G), the role of big pharmaceutical companies (BigPharma) and Bill Gates (BillGates) in the spread of COVID-19, the claim that the scope of the pandemic was exaggerated (Exaggeration), the claim that hospitals were actually empty (FilmYourHospital), the idea that the genetically modified crops led to the emergence of COVID-19 (GMO), and the skepticism regarding the ineffectiveness and side-effects of vaccines (Vaccines).

We combine the conspiracy-specific terms adopted by prior studies \cite{erokhin_covid-19_2022,moffitt_hunting_2021} and compile an extensive list of keywords and phrases to identify the eight conspiracy theories (Table \ref{tab:conspiracy_keywords}). Subsequently, we match these keywords with the lemmatized words in each news claim or post, and for phrases, we directly match them with the original translated text. We find that \num{89} false news claims and \num{4265} false news posts incorporate keywords (phrases) indicating conspiracy theories. Given that the shares of conspiracy posts range from 0.6\% to 18\% in previous studies \cite{erokhin_covid-19_2022}, the 6\% share of posts containing conspiracy theories in our dataset is reasonable. Upon a manual check of the \num{89} source news items that contain keywords or phrases related to conspiracy theories, we find that the accuracy is 92.1\%. Therefore, the selected keywords (phrases) can effectively identify conspiracy theories in our dataset.

\begin{table}
\centering
\caption{COVID-19 conspiracy theories and corresponding keywords (phrases). ``()'' indicates that the word can be separated into a phrase. ``\{\} and \{\}'' indicates one of the words in the first set \{\} needs to appear with one of the words in the second set \{\}. ``5G'' and ``Bill Gates'' are case sensitive.}
\begin{tabularx}{\columnwidth}{@{\hspace{\tabcolsep}\extracolsep{\fill}}ll}
\toprule
\multicolumn{1}{l}{\textbf{Conspiracy theory}}&\multicolumn{1}{l}{\textbf{Keywords}}\\
\midrule
Origin  &bat, batsoup, weapon, biowar, lab, conspiracy, chinese()virus, \\
&biochemical, bioterrorism, biowarfare, china()virus, wildlife, ccp()virus,\\
&chinese()flu, china()coronavirus, chinese()coronavirus, chinesevirus19,\\
&chines()virus, manmade, wuhan()coronavirus, wuhan()coronavius, \\
&wuflu,wuhan()virus, chinaliedpeopledie\\
\rowcolor{gray!50}
Vaccines    &\{vaccine, vax, vaccination, vaccin, vacine\} and \{anti, infertile, \\
\rowcolor{gray!50}
&do not work, don't work, does not work, doesn't work, autism, \\
\rowcolor{gray!50}
&autoimmune\}\\
5G      &5G\\
\rowcolor{gray!50}
BigPharma   &big pharma, fauci pharma, gates pharma\\
BillGates   &Bill Gates\\
\rowcolor{gray!50}
Exaggeration    &does not exist, doesn't exist, exaggerated, inflated, refuse\\
FilmYourHospital    &filmyourhospital, film your hospital, empty hospital, empty bed\\
\rowcolor{gray!50}
GMO &gmo, genetically modified\\
\bottomrule
\end{tabularx}
\label{tab:conspiracy_keywords}
\end{table}

\subsection{User Engagement on X}

\begin{figure}
\centering
\includegraphics[width=\textwidth]{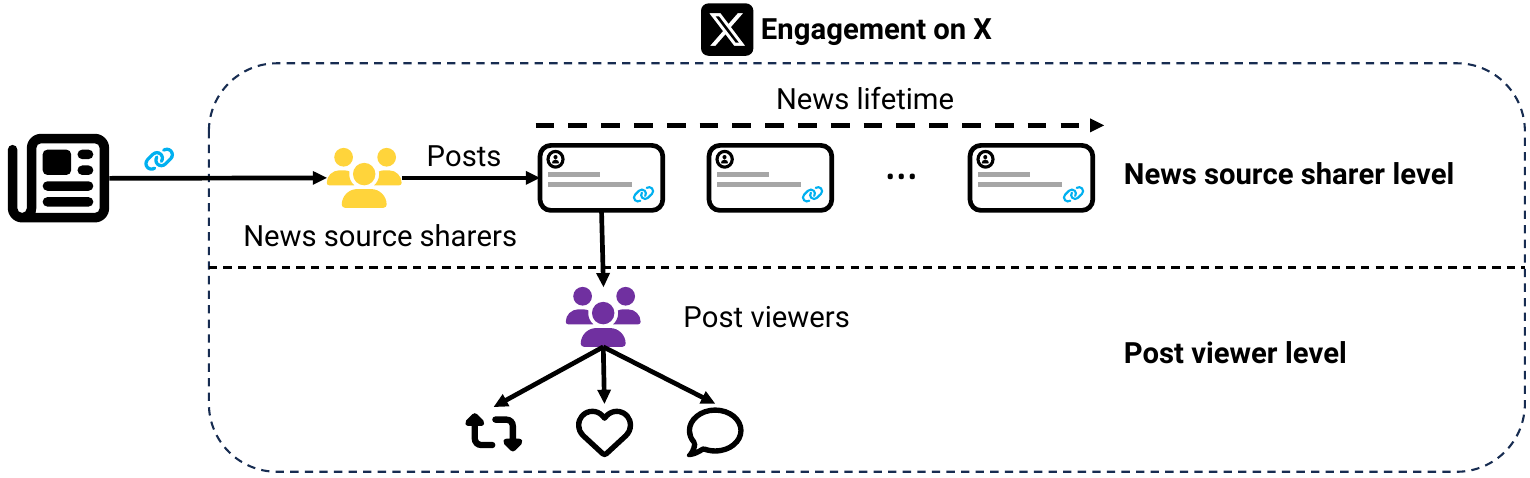}
\vspace{1pt}
\caption{Engagement on X at news source sharer level and post viewer level. The news source sharers view original news items and share them on X through posts containing news links. Subsequently, post viewers engage with these news posts by reposting, liking, or replying.}
\label{fig:engagement_levels}
\end{figure}

On the platform of X, users can interact with posts they view through reposting, liking, or replying. Existing research typically uses the number of reposts to study engagement with misinformation on social media \cite{vosoughi_spread_2018,chuai_anger_2022,prollochs_emotions_2021,drolsbach_diffusion_2023,chuai_roll-out_2023}. Additionally, the number of likes and the number of replies are also important indicators of engagement with online posts \cite{chuai_roll-out_2023,papakyriakopoulos_impact_2022}. However, few studies examine engagement with misinformation from the perspective of news source sharing. Multiple posts on X can link to the same source news items, which increases the visibility of the news items. In this paper, we consider user engagement at two levels: news source sharer level and post viewer level (Fig. \ref{fig:engagement_levels}). At news source sharer level, news items are shared by news source sharers via posts on X. The engagement metrics include the number of posts and the lifetime of news items from the creation of the first post to the creation of the latest post. At post viewer level, engagement is measured by how viewers interact with these posts. Post viewers can repost, like, and reply to posts that are created by news source sharers and contain links to the original news items.

\subsection{Empirical Models}

We employ multiple explanatory regression models to explore our research questions separately. To this end, we define the following variables to measure \emph{engagement at news source sharer level and post viewer level}.
\begin{itemize}[leftmargin=*]
    \item The engagement at news source sharer level:
    \begin{itemize}
        \item $\var{NewsLifetime}$: The duration (in hours) from the creation of the first post to the creation of the latest post on X for the specific news item.
        \item $\var{NewsPostCount}$: A count variable indicating the number of posts associated with the specific news item.
    \end{itemize}
    \item The engagement at post viewer level:
    \begin{itemize}
        \item $\var{RepostCount}$: A count variable indicating the number of reposts received by the specific post.
        \item $\var{LikeCount}$: A count variable indicating the number of likes received by the specific post.
        \item $\var{ReplyCount}$: A count variable indicating the number of replies received by the specific post.
    \end{itemize}
\end{itemize}
Additionally, we define the following \emph{key variables} to address our research questions:
\begin{itemize}[leftmargin=*]
    \item $\var{Falsehood}$: A dummy variable indicating whether the news claim is false ($=$ 1) or true ($=$ 0).
    \item $\var{Conspiracy}$: A dummy variable indicating whether the false news claim or post contains conspiracy theories ($=$ 1) or not ($=$ 0).
    \item $\var{Diversity}$: The topic diversity of the news claim or post.
\end{itemize}
We also control for the variables from \emph{content characteristics} that can affect user engagement:
\begin{itemize}[leftmargin=*]
    \item $\var{Words}$: A continuous-like count variable indicating the number of words in the text of the news claim or post. This control variable is commonly used to account for the content richness \cite{miani_interconnectedness_2022}.
    \item $\var{Media}$: A dummy variable indicating whether the post has media elements ($=$ 1) or not ($=$ 0). Previous research has found that media elements, such as images and videos, can facilitate user engagement \cite{zhou_linguistic_2021}.
    \item Emotions are considered to be the main driver of online information diffusion \cite{chuai_anger_2022,prollochs_emotions_2021,prollochs_emotions_2021-1}. Our analysis considers six cross-cultural emotions -- anger, disgust, fear, joy, sadness, and surprise -- out of the eight basic emotions \cite{chuai_anger_2022,plutchik_emotions_1991}. We employ the commonly used NRC emotion lexicon to count emotion words in news claims and posts. Notably, 92.1\% news claims and 95\% posts contain at least one emotion word. Subsequently, we calculate the weight of each emotion in the specific text content based on the number of corresponding emotion words. If the news claims (posts) have no emotion words, all the weights of emotions are set to 0.
\end{itemize}
Finally, the posters on social media play an important role in driving engagement with their posts \cite{vosoughi_spread_2018,chuai_anger_2022}. We use the following four variables to account for \emph{poster characteristics}: 
\begin{itemize}[leftmargin=*]
    \item $\var{Verified}$: A dummy variable indicating whether the post account is verified ($=$ 1) or not ($=$ 0).
    \item $\var{AccountAge}$: A continuous-like count variable indicating the age (in days) of the post account.
    \item $\var{Followers}$: A continuous-like count variable indicating the number of followers of the post account.
    \item $\var{Followees}$: A continuous-like count variable indicating the number of followees of the post account.
\end{itemize}
The descriptive statistics for the above variables are shown in Table \ref{tab:data_summary}.

\begin{table}
\centering
\caption{Dataset overview. Reported are mean values or count numbers for the variables (standard deviations in parentheses). Column (1) includes all the news claims and associated posts. 51.7\% in English, 21.6\% in Spanish, 4.9\% in Portuguese, 15.7\% in Hindi, 3.2\% in French and 3.0\% in Italian. Column (2) includes false news claims (posts) that contain conspiracy theories.}
\begin{tabularx}{\columnwidth}{@{\hspace{\tabcolsep}\extracolsep{\fill}}l*{2}{c}}
\toprule
& (1) & (2)\\
& All & CTs\\
\midrule
News Count & {7,273} & {82} \\
Falsehood & {18.9\%} & {100\%}\\
\multicolumn{3}{l}{\underline{Engagement metrics}}\\
\quad Lifetime & 315.617 \small (900.617) & 1,968.141 \small (1,626.337) \\
\quad Post Count & {70,904} & {4,265} \\
\quad Repost Count & {1,300,597} & {161,823} \\
\quad Like Count & {3,943,848} & {452,106} \\
\quad Reply Count & {380,224} & {107,363} \\
\multicolumn{3}{l}{\underline{News characteristics}}\\
\quad Dates & {01/02/20~--~09/14/20} & {01/07/20~--~09/08/20} \\
\quad Diversity & 0.083 \small (0.045) & 0.089 \small (0.038) \\
\quad Words & 15.614 \small (6.364) & 11.841 \small (6.739) \\
\quad Anger & 0.058 \small (0.109) & 0.040 \small (0.093) \\
\quad Disgust & 0.045 \small (0.108) & 0.059 \small (0.153) \\
\quad Fear & 0.183 \small (0.208) & 0.149 \small (0.183) \\
\quad Joy & 0.065 \small (0.128) & 0.097 \small (0.215) \\
\quad Sadness & 0.096 \small (0.139) & 0.089 \small (0.135) \\
\quad Surprise & 0.045 \small (0.099) & 0.026 \small (0.080) \\
\multicolumn{3}{l}{\underline{Post characteristics}}\\
\quad Dates & {01/02/20~--~09/14/20} & {01/09/20~--~09/09/20} \\
\quad Diversity & 0.104 \small (0.052) & 0.111 \small (0.050) \\
\quad Words & 26.372 \small (10.192) & 27.006 \small (8.876) \\
\quad Anger & 0.068 \small (0.100) & 0.069 \small (0.090) \\
\quad Disgust & 0.074 \small (0.134) & 0.064 \small (0.105) \\
\quad Fear & 0.195 \small (0.177) & 0.157 \small (0.146) \\
\quad Joy & 0.062 \small (0.106) & 0.088 \small (0.163) \\
\quad Sadness & 0.129 \small (0.133) & 0.095 \small (0.106) \\
\quad Surprise & 0.052 \small (0.091) & 0.050 \small (0.088) \\
\quad Media & {21.9\%} & {13.6\%} \\
\quad Verified & {18.5\%} & {7.4\%} \\
\quad AccountAge & 2,460.659 \small (1,457.714) & 2,259.347 \small (1,393.434) \\
\quad Followers & 369,041.439 \small (1,755,545.224) & 56,847.547 \small (1,399,857.795) \\
\quad Followees & 3,031.517 \small (15,099.135) & 2,361.965 \small (15,535.361) \\
\bottomrule
\end{tabularx}
\label{tab:data_summary}
\end{table}

\textbf{Topic diversity (RQ1.1 \& RQ2.1):} 
We use $\var{Diversity}$ as the dependent variable to analyze topic diversity in true and false news from original news claims to online posts. In terms of claims, the linear regression model that incorporates the main independent variables of $\var{Falsehood}$ and $\var{Conspiracy}$ is specified as: 
\begin{equation}
\begin{aligned}
    \var{Diversity_{i}}= \, \beta_{0} + \beta_{1}\var{Falsehood_{i}} + \beta_{2}\var{Conspiracy_{i}} + \beta_{3}\var{Words_{i}} + \bm{\alpha}^{'}\bm{Emotions_{i}} + u_{\text{lang}} + u_{\text{time}},
\end{aligned}
\label{equ:diversity_news}
\end{equation}
where $\bm{\var{Emotions}}$ indicates six basic emotions. Additionally, $u_{\text{lang}}$ denotes language-specific fixed effects, and $u_{\text{time}}$ denotes month-year fixed effects.

Given that each news item can have multiple posts, we need to control for the possible heterogeneity among news items when analyzing the topic diversity in terms of online posts. Therefore, we specify a liner regression model incorporating news-specific random effects:
\begin{equation}
\begin{aligned}
\var{Diversity_{i}}= \, &  \beta_{0} + \beta_{1}\var{Falsehood_{i}} + \beta_{2}\var{Conspiracy_{i}} + \beta_{3}\var{Words_{i}} + \beta_{4}\var{Media_{i}} + \beta_{5}\var{Verified_{i}} \\
&+ \beta_{6}\var{AccountAge_{i}} + \beta_{7}\var{Followers_{i}} + \beta_{8}\var{Followees_{i}} 
     + \bm{\alpha}^{'}\bm{Emotions_{i}} \\
&+ u_{\text{lang}} + u_{\text{time}} + v_{\text{news}},
\end{aligned}
\label{equ:diversity_post}
\end{equation}
where $v_{\text{news}}$ indicates news-specific random effects. Additionally, we incorporate $\var{Media}$, poster characteristics as control variables. In the above two models, the coefficient estimate of $\var{Falsehood}$ ($\beta_{1}$) is to address RQ1.1, and the coefficient estimate of $\var{Conspiracy}$ ($\beta_{2}$) is to address RQ2.1.

\textbf{Engagement at news source sharer level -- news lifetime and post count (RQ1.2 \& RQ2.2):}
The first layer of engagement is from source news claims to news posts on X. We examine how topic diversity and conspiracy theories influence engagement with misinformation in terms of news lifetime and post count. First, we take $\var{NewsLifetime}$ as the dependent variable and specify a linear regression model as follows:
\begin{equation}
\begin{aligned}
    \var{Lifetime_{i}}= \, & \beta_{0} + \beta_{1}\var{Falsehood_{i}} + \beta_{2}\var{Conspiracy_{i}} 
    + \beta_{3}\var{Diversity_{i}} + \beta_{4}\var{Falsehood_{i}} \times \var{Diversity_{i}} \\
    &+ \beta_{4}\var{Words_{i}} + \bm{\alpha}^{'}\bm{Emotions_{i}} + u_{\text{lang}} + u_{\text{time}},
\end{aligned}
\label{equ:lifetime}
\end{equation}
where the interaction term of $\var{Falsehood} \times \var{Diversity}$ is to estimate the moderating effect of $\var{Falsehood}$ on $\var{Diversity}$. Subsequently, we use a negative binomial regression model to explain $\var{PostCount}$. The dependent variable is $\var{log(E(PostCount_{i}|\bm{x_{i}}))}$, and the independent variables remain consistent with Eq. (\ref{equ:lifetime}).

\textbf{Engagement at post viewer level -- reposts, likes, and replies (RQ1.2 \& RQ2.2):}
To examine the effects of topic diversity and conspiracy theories on social engagement (\ie, reposts, likes, and replies) with misinformation on X, we specify a negative binomial regression model with language-specific fixed effects, month-year fixed effects, and news-specific random effects:
\begin{equation}
\begin{aligned}
     \var{log(E(y_{i}|\bm{x_{i}}))} = \, & \beta_{0} + \beta_{1}\var{Falsehood_{i}} 
     + \beta_{2}\var{Conspiracy_{i}} + \beta_{3}\var{Diversity_{i}} 
     + \beta_{4}\var{Falsehood_{i}} \times \var{Diversity_{i}}\\
     &+ \beta_{3}\var{Words_{i}} 
     + \beta_{4}\var{Media_{i}} + \beta_{5}\var{Verified_{i}} + \beta_{6}\var{AccountAge_{i}} 
     + \beta_{7}\var{Followers_{i}}  \\
     &+ \beta_{8}\var{Followees_{i}} + \bm{\alpha}^{'}\bm{Emotions_{i}} 
     + u_{\text{lang}} + u_{\text{time}} + v_{\text{news}},
\end{aligned}
\end{equation}
where $\var{x_{i}}$ indicates all the independent variables, and $\var{y_{i}}$ signifies the dependent variables, including $\var{RepostCount_{i}}$, $\var{LikeCount_{i}}$ and $\var{ReplyCount_{i}}$. In the above models for engagement, the coefficient estimates of $\var{Falsehood}$ ($\beta_{1}$), $\var{Diversity}$ ($\beta_{3}$), and $\var{Falsehood} \times \var{Diversity}$ ($\beta_{4}$) are to address RQ1.2, and the coefficient estimate of $\var{Conspiracy}$ ($\beta_{2}$) is to address RQ2.2. All the continuous variables in these models are z-standardized.

\section{Empirical Results}

\subsection{Topic Diversity (RQ1.1 \& RQ2.1)}
\label{sec:reg_topic_diversity}

\begin{figure}
\centering
\begin{subfigure}{0.32\textwidth}
\caption{}
\includegraphics[width=\textwidth]{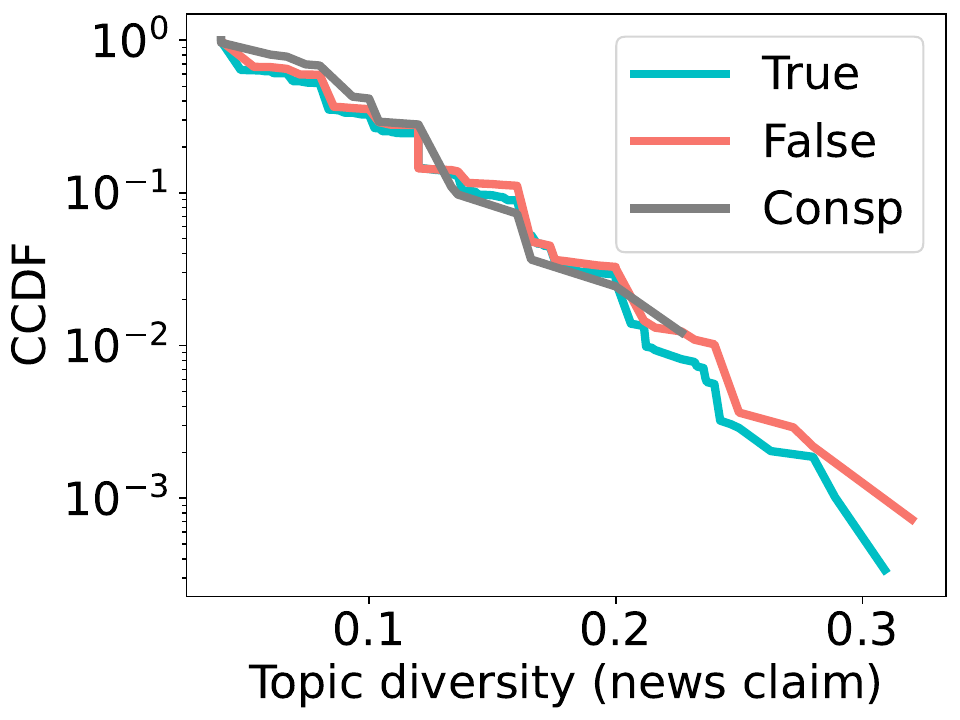}
\label{fig:ccdf_topic_gini_claim}
\end{subfigure}
\hfill
\begin{subfigure}{0.32\textwidth}
\caption{}
\includegraphics[width=\textwidth]{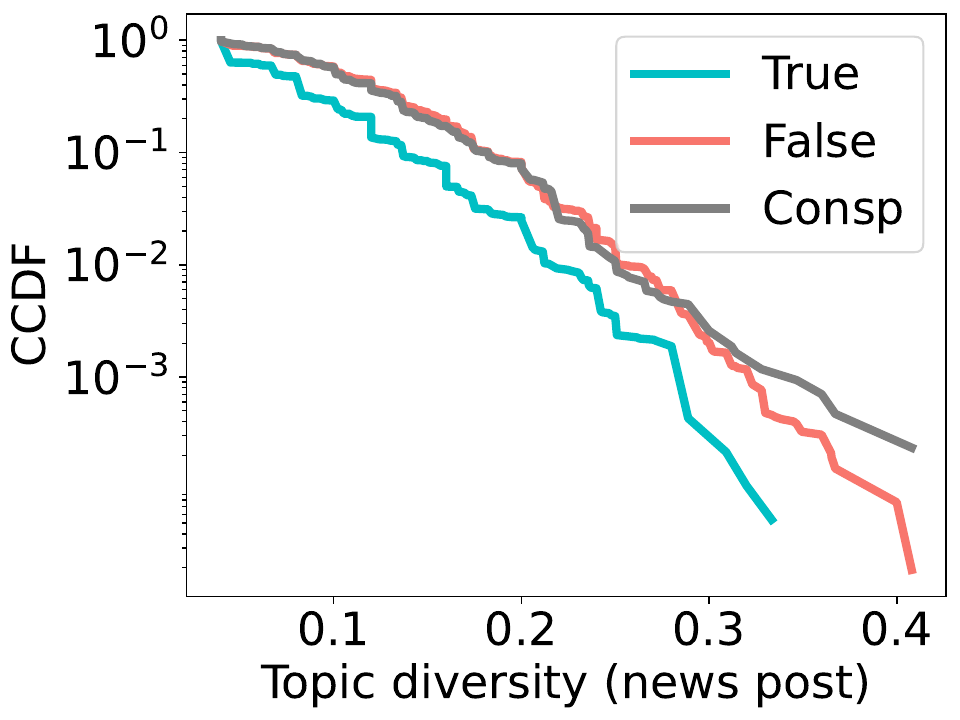}
\label{fig:ccdf_topic_gini_tweet}
\end{subfigure}
\hfill
\begin{subfigure}{0.32\textwidth}
\caption{}
\includegraphics[width=\textwidth]{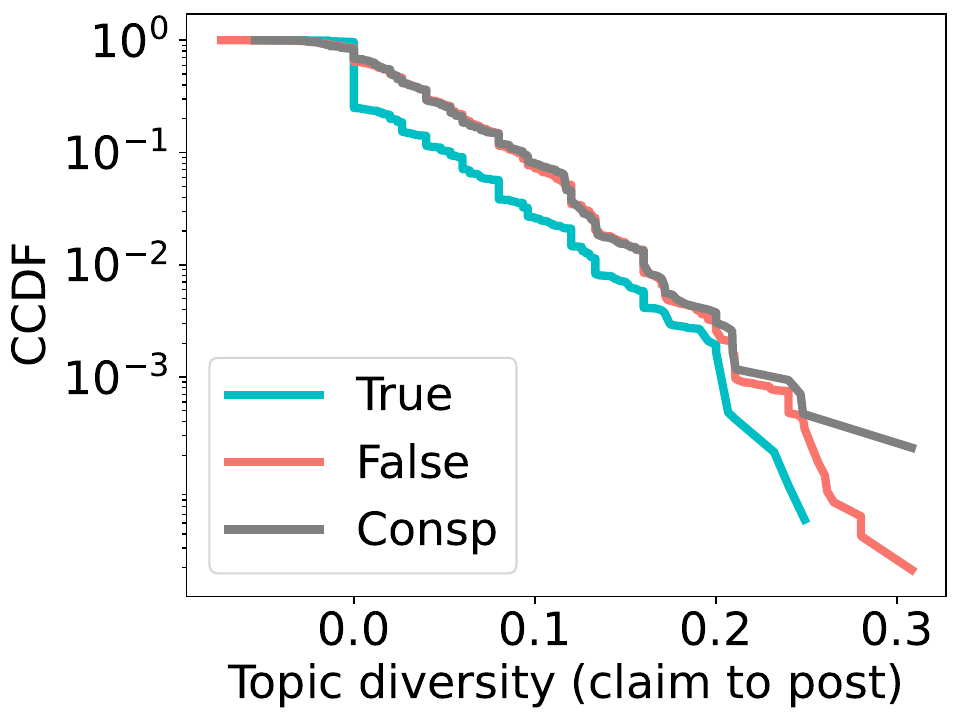}
\label{fig:ccdf_topic_gini_diff}
\end{subfigure}

\begin{subfigure}{0.32\textwidth}
\caption{}
\includegraphics[width=\textwidth]{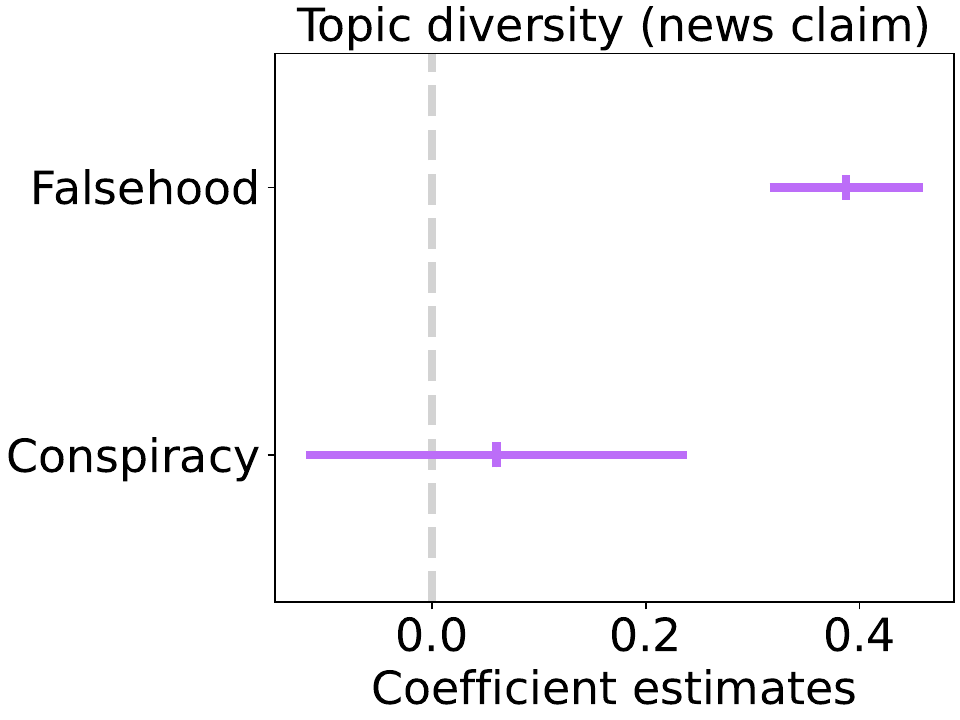}
\label{fig:topic_diversity_news_claim_coefs}
\end{subfigure}
\hfill
\begin{subfigure}{0.32\textwidth}
\caption{}
\includegraphics[width=\textwidth]{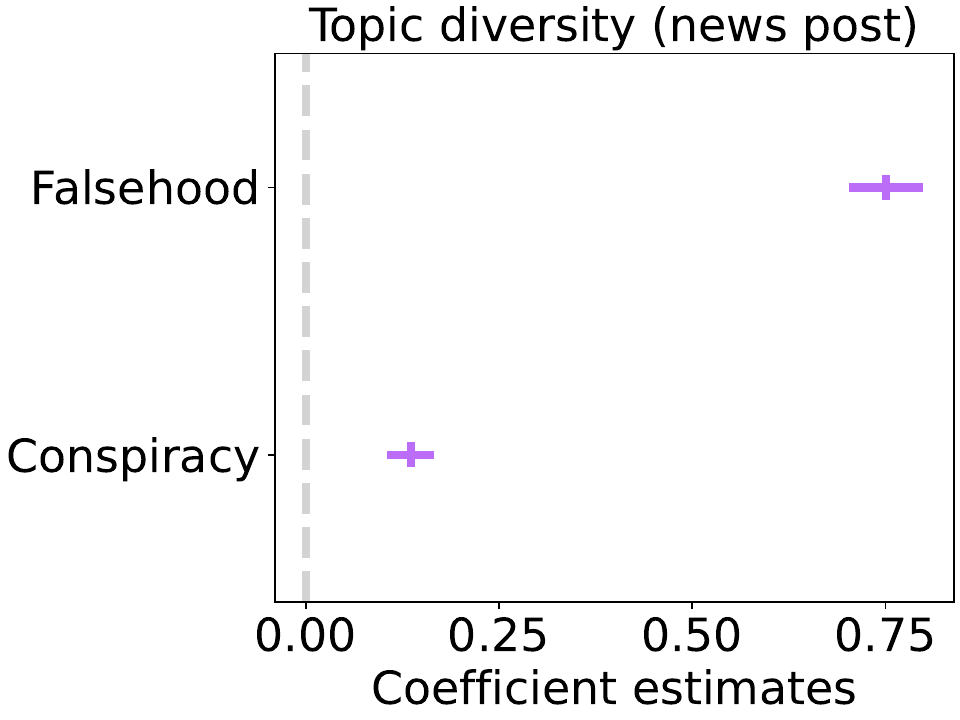}
\label{fig:topic_diversity_news_post_coefs}
\end{subfigure}
\hfill
\begin{subfigure}{0.32\textwidth}
\caption{}
\includegraphics[width=\textwidth]{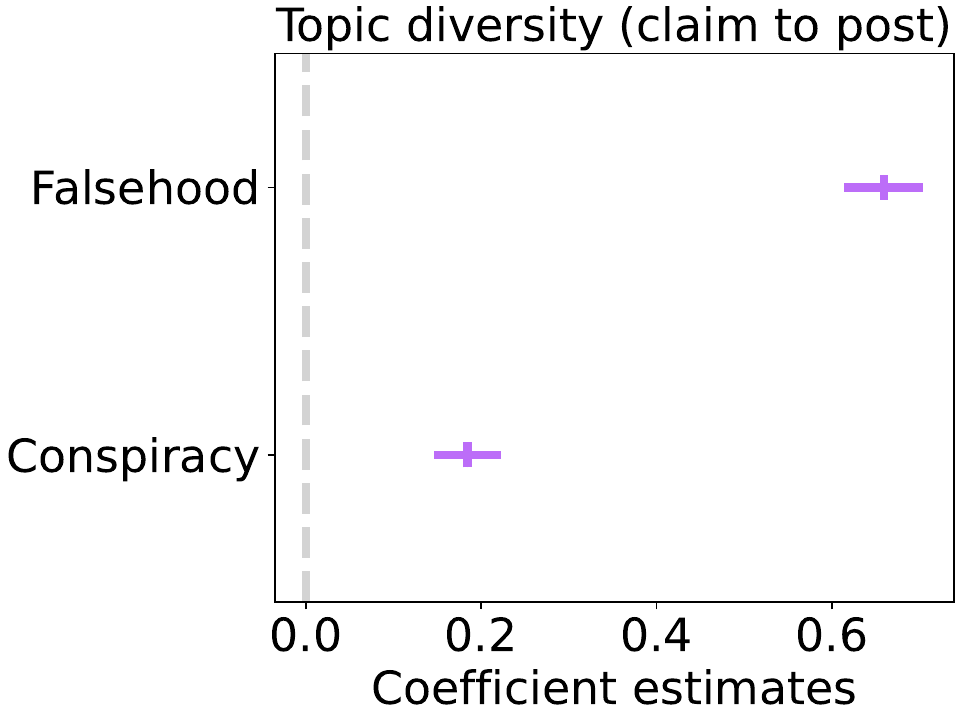}
\label{fig:topic_diversity_claim_to_post_coefs}
\end{subfigure}
\caption{The topic diversity from news claims to news posts. \subref{fig:ccdf_topic_gini_claim}--\subref{fig:ccdf_topic_gini_diff} The Complementary Cumulative Distribution Functions (CCDFs) for \subref{fig:ccdf_topic_gini_claim} topic diversity at the source claim level, \subref{fig:ccdf_topic_gini_tweet} topic diversity at the post level, and \subref{fig:ccdf_topic_gini_diff} the changes in topic diversity from claims to posts. Three categories of CCDFs are included in each figure, \ie, true news (True), false news (False), and CT news (Consp). \subref{fig:topic_diversity_news_claim_coefs}--\subref{fig:topic_diversity_claim_to_post_coefs} The coefficient estimates of the independent variables -- $\var{Falsehood}$ and $\var{Conspiracy}$. Other independent variables are included during estimation but omitted in the visualization for better readability. Shown are mean values with error bars representing 95\% Confidence Intervals (CIs). The dependent variables are \subref{fig:topic_diversity_news_claim_coefs} topic diversity in news claims, \subref{fig:topic_diversity_news_post_coefs} topic diversity in news posts, and \subref{fig:topic_diversity_claim_to_post_coefs} the changes in topic diversity from claims to posts, respectively. See Table \ref{tab:topic_diversity} in Suppl. \ref{sec:main_estimation} for full estimation results.}
\label{fig:topic_diversity}
\end{figure}

\subsubsection{Descriptive statistics.}
We start by examining topic diversity from news claims to news posts (Figs. \ref{fig:ccdf_topic_gini_claim}--\ref{fig:ccdf_topic_gini_diff}). Fig. \ref{fig:ccdf_topic_gini_claim} shows that false news claims (mean of 0.086) exhibit 4.9\% higher topic diversity compared to true news claims (mean of 0.082, $\var{KS} =$ \num{0.072}, $p < $ \num{0.001}). This difference is significantly magnified (43.0\% higher) at the post level, where the average topic diversity in false news posts is 0.113, compared to 0.079 in true news posts (Fig. \ref{fig:ccdf_topic_gini_tweet}, $\var{KS} =$ \num{0.333}, $p < $ \num{0.001}). Additionally, we examine the paired changes in topic diversity from claims to their corresponding posts (Fig. \ref{fig:ccdf_topic_gini_diff}), finding that topic diversity in false news items (increase of 0.031) increases 158.3\% more than that in true news items (increase of 0.012, $\var{KS} =$ \num{0.396}, $p < $ \num{0.001}). For false news items containing conspiracy theories (CTs), the topic diversity at the claim level has no statistically significant difference compared to the overall false news items (Fig. \ref{fig:ccdf_topic_gini_claim}, $\var{KS} =$ \num{0.138}, $p = $ \num{0.095}). However, Fig. \ref{fig:ccdf_topic_gini_diff} indicates that the most significant increases in topic diversity from claims to posts are more likely to be in CT news items, compared to overall false news items ($\var{KS} =$ \num{0.044}, $p < $ \num{0.001}). This may lead to the significant difference in topic diversity between CT news and general false news at the post level  (Fig. \ref{fig:ccdf_topic_gini_tweet}, $\var{KS} =$ \num{0.036}, $p < $ \num{0.001}).

\subsubsection{Regression results.}
\label{sec:topic_diversity_regression}
We further conduct regression analysis to estimate the effects of $\var{Falsehood}$ and $\var{Conspiracy}$ on topic diversity (Figs. \ref{fig:topic_diversity_news_claim_coefs}--\ref{fig:topic_diversity_claim_to_post_coefs}). In Fig. \ref{fig:topic_diversity_news_claim_coefs}, the coefficient estimate of $\var{Falsehood}$ is significantly positive ($\var{coef.} =$ \num{0.388}, $\var{p} <$ \num{0.001}). This implies that the topic diversity in false news is, on average, 0.388 standard deviations higher than in true news at the claim level, consistent with the observations in Fig. \ref{fig:ccdf_topic_gini_claim}. Additionally, the coefficient estimate of $\var{Conspiracy}$ is not statistically significant, which indicates that the topic diversity in false news claims containing conspiracy theories is not significantly different from that in false news claims without conspiracy theories (Fig. \ref{fig:topic_diversity_news_claim_coefs}). Subsequently, we explore the topic diversity between true and false news at the post level. In Fig. \ref{fig:topic_diversity_news_post_coefs}, the coefficient estimate of $\var{Falsehood}$ is significantly positive ($\var{coef.} =$ \num{0.751}, $\var{p} <$ \num{0.001}), suggesting that the topic diversity in false news is, on average, 0.751 standard deviations higher than in true news at the post level.

Notably, as observed in the CCDF in Fig. \ref{fig:ccdf_topic_gini_tweet} and the coefficient estimates of $\var{Falsehood}$ in Figs. \ref{fig:topic_diversity_news_claim_coefs}--\ref{fig:topic_diversity_news_post_coefs}, the difference between true and false news posts is larger than that between true and false news claims. Additionally, the coefficient estimate of $\var{Conspiracy}$ in Fig. \ref{fig:topic_diversity_news_post_coefs} becomes significantly positive ($\var{coef.} =$ \num{0.136}, $\var{p} <$ \num{0.001}). This suggests that false news posts containing conspiracy theories exhibit even higher topic diversity (0.136 standard deviations more) than false news posts without conspiracy theories. There are two possible reasons for the larger difference in topic diversity between true and false news at the post level compared to the claim level. One is that news claims with higher topic diversity receive more posts on X. Another is that false news has a bigger increase in topic diversity from the claim level to the post level, compared to true news. Given the second reason, we further examine the paired changes in topic diversity from the source news claims to the corresponding posts. In Fig. \ref{fig:topic_diversity_claim_to_post_coefs}, the positive coefficient estimate of $\var{Falsehood}$ is statistically significant ($\var{coef.} =$ \num{0.659}, $\var{p} <$ \num{0.001}). This suggests that the increase in topic diversity in false news from the claim level to the post level is significantly larger (0.659 standard deviations more) than in true news. This finding could partially explain the enlarged difference in topic diversity between true and false news posts, compared to their source claims. In addition, the coefficient estimate of $\var{Conspiracy}$ in Fig. \ref{fig:topic_diversity_claim_to_post_coefs} is significantly positive ($\var{coef.} =$ \num{0.185}, $\var{p} <$ \num{0.001}). This means that the topic diversity in false news that contains conspiracy theories increases 0.185 standard deviations more than that in false news without conspiracy theories from the claim level to the post level. This finding could partially explain why the coefficient estimate of $\var{Conspiracy}$ in Fig. \ref{fig:topic_diversity_news_claim_coefs} is not statistically significant, whereas it becomes statistically significant in Fig. \ref{fig:topic_diversity_news_post_coefs}.

In summary, we find that false news claims exhibit higher topic diversity compared to true news claims, and this difference is magnified at the post level (RQ1.1). Additionally, false news posts containing conspiracy theories have higher topic diversity than those without conspiracy theories, while no significant difference is observed at the claim level (RQ2.1).

\subsection{News Source Sharer Engagement: News Lifetime and Post Count (RQ1.2 \& RQ2.2)}

\begin{figure}
\centering
\begin{subfigure}{0.32\textwidth}
\caption{}
\includegraphics[width=\textwidth]{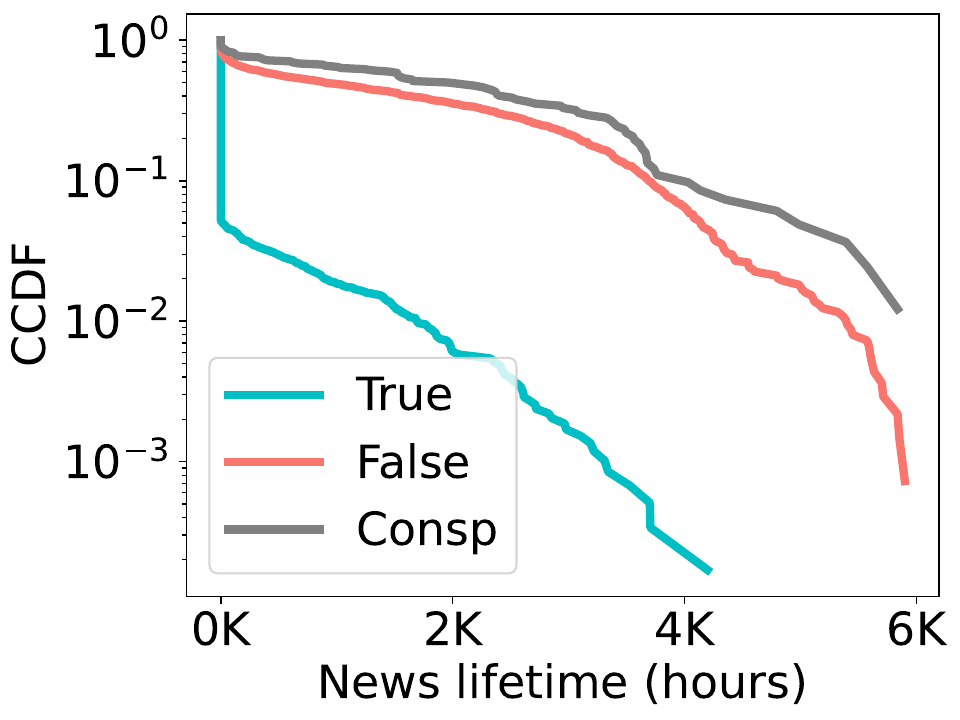}
\label{fig:ccdf_lifetime}
\end{subfigure}
\hspace{20pt}
\begin{subfigure}{0.32\textwidth}
\caption{}
\includegraphics[width=\textwidth]{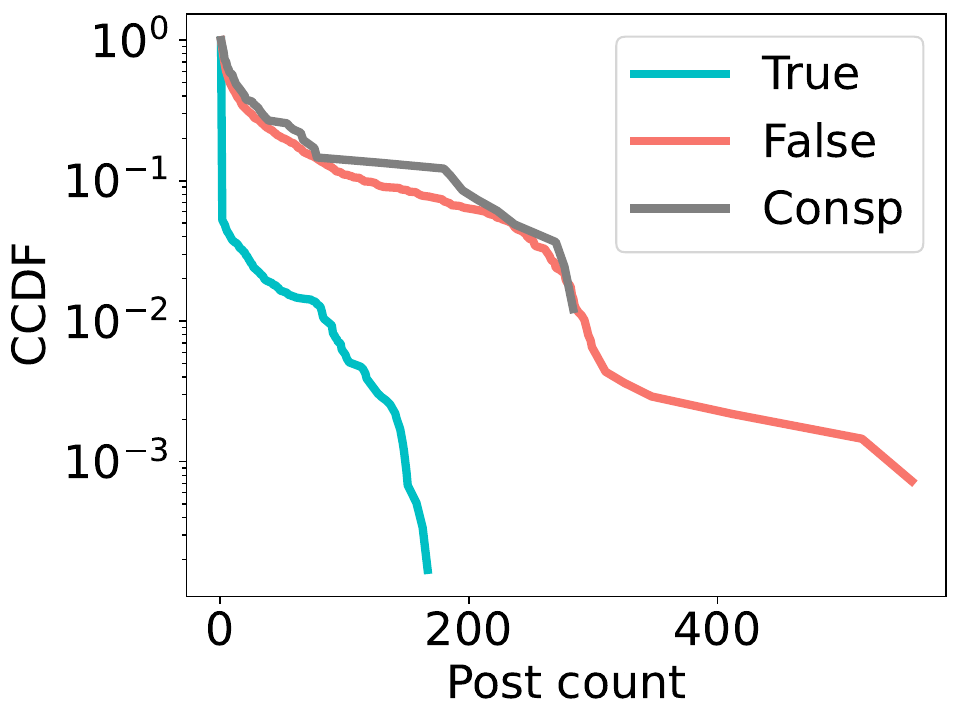}
\label{fig:ccdf_tweet_num}
\end{subfigure}

\begin{subfigure}{0.32\textwidth}
\caption{}
\includegraphics[width=\textwidth]{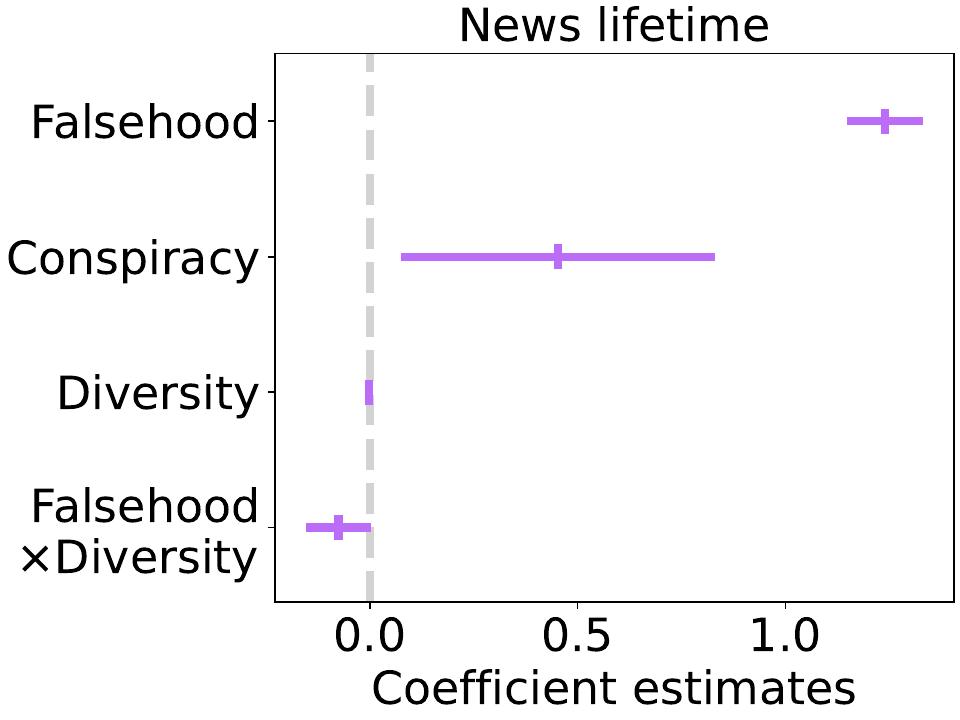}
\label{fig:news_lifetime_coefs}
\end{subfigure}
\hspace{20pt}
\begin{subfigure}{0.32\textwidth}
\caption{}
\includegraphics[width=\textwidth]{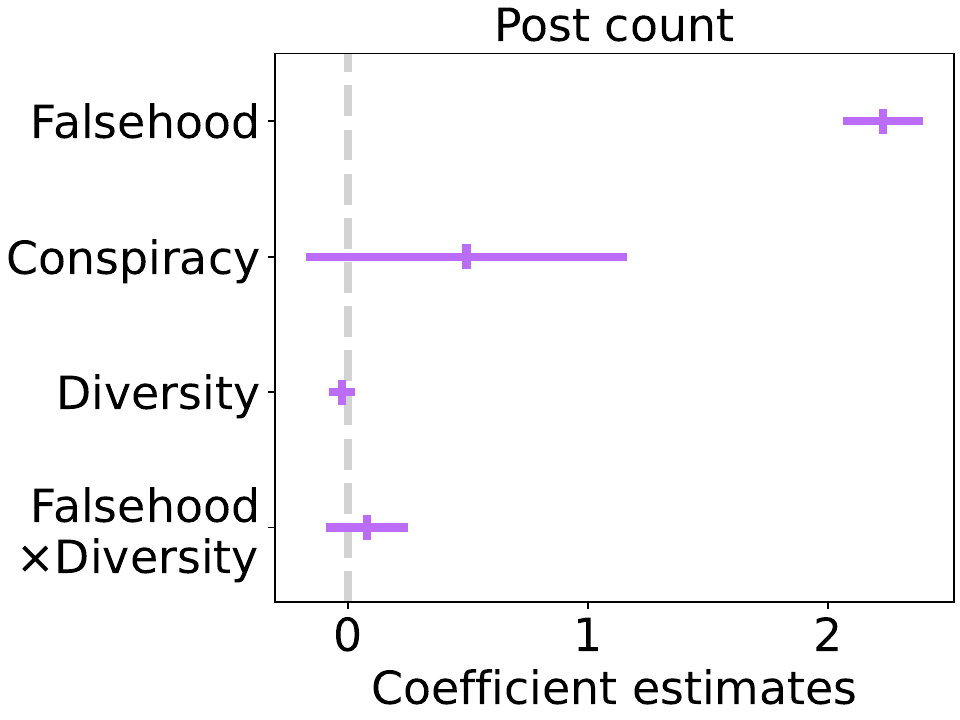}
\label{fig:post_count_coefs}
\end{subfigure}
\caption{The engagement at news source sharer level -- news lifetime and post count. \subref{fig:ccdf_lifetime}--\subref{fig:ccdf_tweet_num} The CCDFs for \subref{fig:ccdf_lifetime} news lifetime and \subref{fig:ccdf_tweet_num} post count. Three categories of CCDFs are included in each figure, \ie, true news (True), false news (False), and CT news (Consp). \subref{fig:news_lifetime_coefs}--\subref{fig:post_count_coefs} The coefficient estimates of the independent variables -- $\var{Falsehood}$, $\var{Conspiracy}$, $\var{Diversity}$, and $\var{Falsehood} \times \var{Diversity}$. Other independent variables are included during estimation but omitted in the visualization for better readability. Shown are mean values with error bars representing 95\% CIs. The dependent variables are \subref{fig:news_lifetime_coefs} news lifetime and \subref{fig:post_count_coefs} post count, respectively. See Table \ref{tab:news2tweets} in Suppl. \ref{sec:main_estimation} for full estimation results.}
\label{fig:lifetime_post_count}
\end{figure}

\subsubsection{Descriptive statistics.}
In terms of engagement at news source sharer level, we examine news lifetime that indicates the duration (in hours) from the creation time of the first post to the creation time of the last post for each news item on X. Fig. \ref{fig:ccdf_lifetime} shows that false news items have significantly longer lifetimes (mean of 1463.757 hours) than true news items (mean of 47.951 hours, $\var{KS} =$ \num{0.805}, $p < $ \num{0.000}). CT news items have longer lifetimes (mean of 1968.141 hours) than false news items ($\var{KS} =$ \num{0.167}, $p < $ \num{0.05}). Additionally, Fig. \ref{fig:ccdf_tweet_num} shows that false news claims receive more posts (mean of 38.015) than true news claims (mean of 3.159, $\var{KS} =$ \num{0.806}, $p < $ \num{0.001}). However, the number of posts received by CT news items (mean of 44.976) does not differ significantly from that received by false news items ($\var{KS} =$ \num{0.113}, $p = $ \num{0.257}).

\subsubsection{Regression results.}
We conduct regression analysis to further rigorously examine the effects of $\var{Falsehood}$, $\var{Conspiracy}$, and $\var{Diversity}$ on news lifetime and post count (Figs. \ref{fig:news_lifetime_coefs}--\ref{fig:post_count_coefs}). In Fig. \ref{fig:news_lifetime_coefs}, the coefficient estimate of $\var{Falsehood}$ is significantly positive ($\var{coef.} =$ \num{1.239}, $\var{p} <$ \num{0.001}). This suggests that, on average, false news items have lifetimes of 1.239 standard deviations longer for receiving posts on X compared to true news items. Additionally, the coefficient estimate of $\var{Conspiracy}$ is also significantly positive ($\var{coef.} =$ \num{0.452}, $\var{p} <$ \num{0.05}). This means that, on average, the lifetimes for false news items containing conspiracy theories are 0.452 standard deviations longer than that for false news items without conspiracy theories. However, the coefficient estimates for $\var{Diversity}$ and $\var{Falsehood} \times \var{Diversity}$ are not statistically significant. This suggests that topic diversity has no significant impact on the lifetimes of news items, irrespective of their veracity.

Additionally, in Fig. \ref{fig:post_count_coefs}, the coefficient estimate of $\var{Falsehood}$ is significantly positive ($\var{coef.} =$ \num{2.230}, $\var{p} <$ \num{0.001}), indicating that the number of posts referencing a specific source news item is significantly associated with its veracity. On average, false news items are linked by $e^{2.230} - 1 = 8.3$ times more posts on X compared to true news items. Notably, the coefficient estimate of $\var{Conspiracy}$ is not statistically significant for post count, which implies that the integration of conspiracy theories has no additional impact on the number of posts received by false news items. The coefficient estimate of $\var{Diversity}$ is not statistically significant. It suggests that the topic diversity of a source news claim has no significant association with the number of online posts it receives. This finding does not support the first reason for the enlarged difference in topic diversity between true and false news posts, compared to that between true and false news claims (see Section \ref{sec:topic_diversity_regression}). Furthermore, the coefficient estimate of the interaction term $\var{Falsehood}$ $\times$ $\var{Diversity}$ is also not statistically significant, reinforcing the finding that topic diversity has no significant effect on post count, irrespective of the veracity of news claims. 

In summary, we find that, at news source sharer level, misinformation with higher topic diversity does not receive more engagement in terms of news lifetime and post count (RQ1.2). However, false news items containing conspiracy theories tend to receive a higher number of posts on X compared to false news items without conspiracy theories but do not have significantly longer lifetimes (RQ2.2).

\subsection{Post Viewer Engagement: Reposts, Likes, and Replies (RQ1.2 \& RQ2.2)}

\begin{figure}
\centering
\begin{subfigure}{0.32\textwidth}
\caption{}
\includegraphics[width=\textwidth]{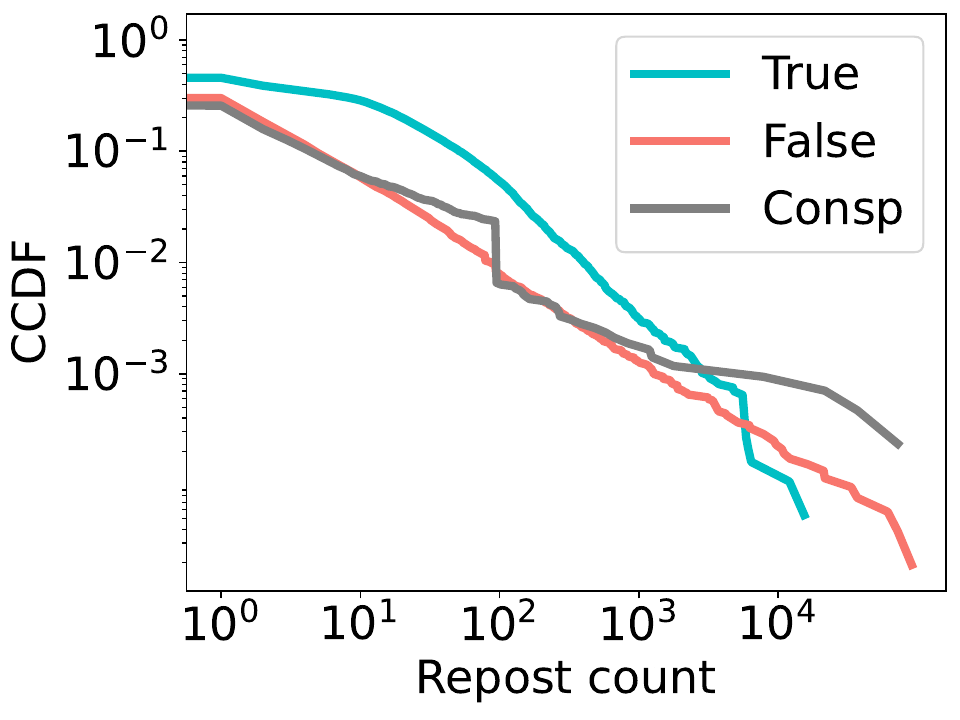}
\label{fig:ccdf_retweet_count}
\end{subfigure}
\hfill
\begin{subfigure}{0.32\textwidth}
\caption{}
\includegraphics[width=\textwidth]{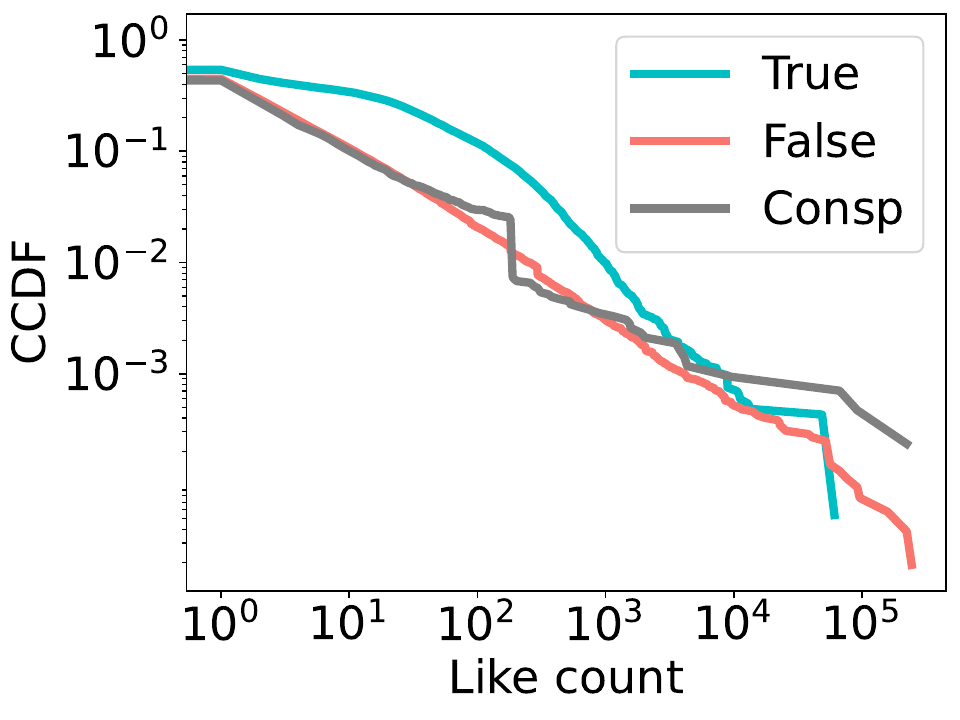}
\label{fig:ccdf_like_count}
\end{subfigure}
\hfill
\begin{subfigure}{0.32\textwidth}
\caption{}
\includegraphics[width=\textwidth]{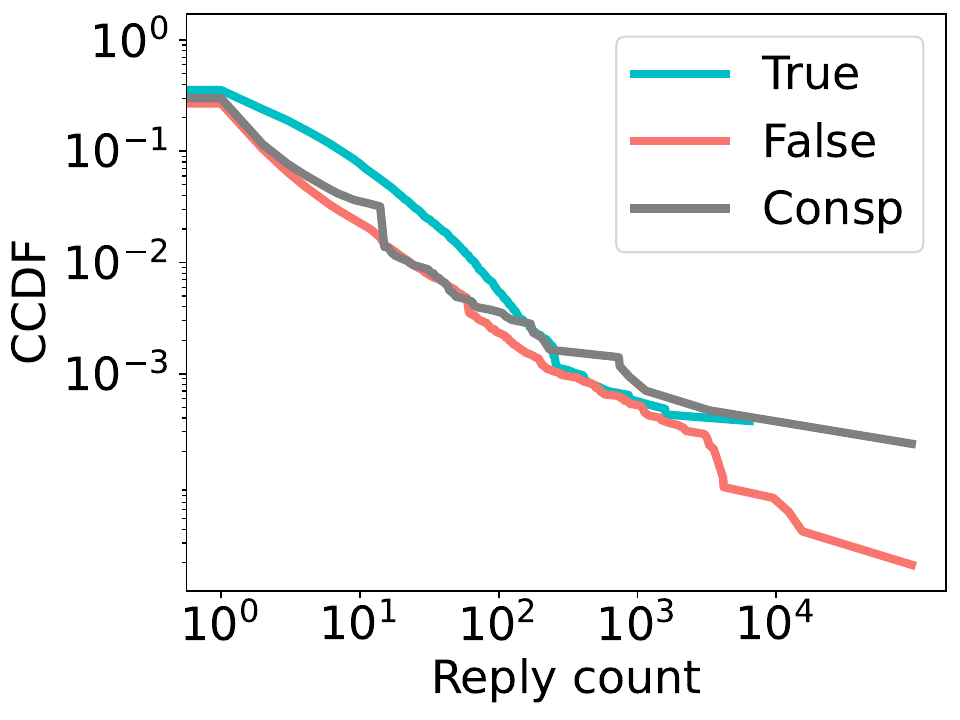}
\label{fig:ccdf_reply_count}
\end{subfigure}

\begin{subfigure}{0.32\textwidth}
\caption{}
\includegraphics[width=\textwidth]{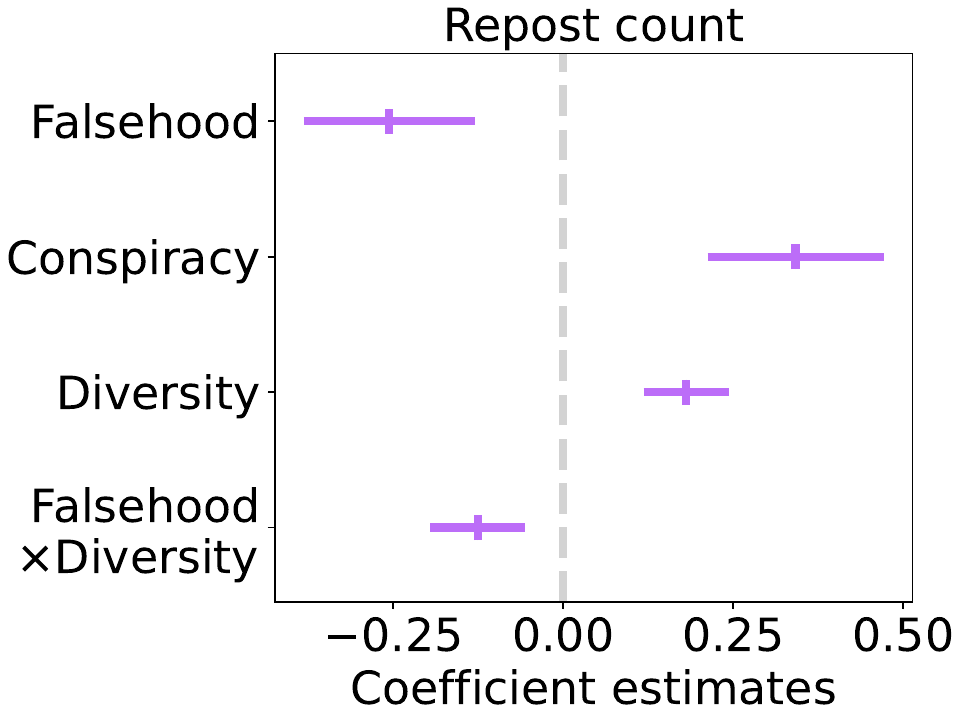}
\label{fig:repost_count_coefs}
\end{subfigure}
\hfill
\begin{subfigure}{0.32\textwidth}
\caption{}
\includegraphics[width=\textwidth]{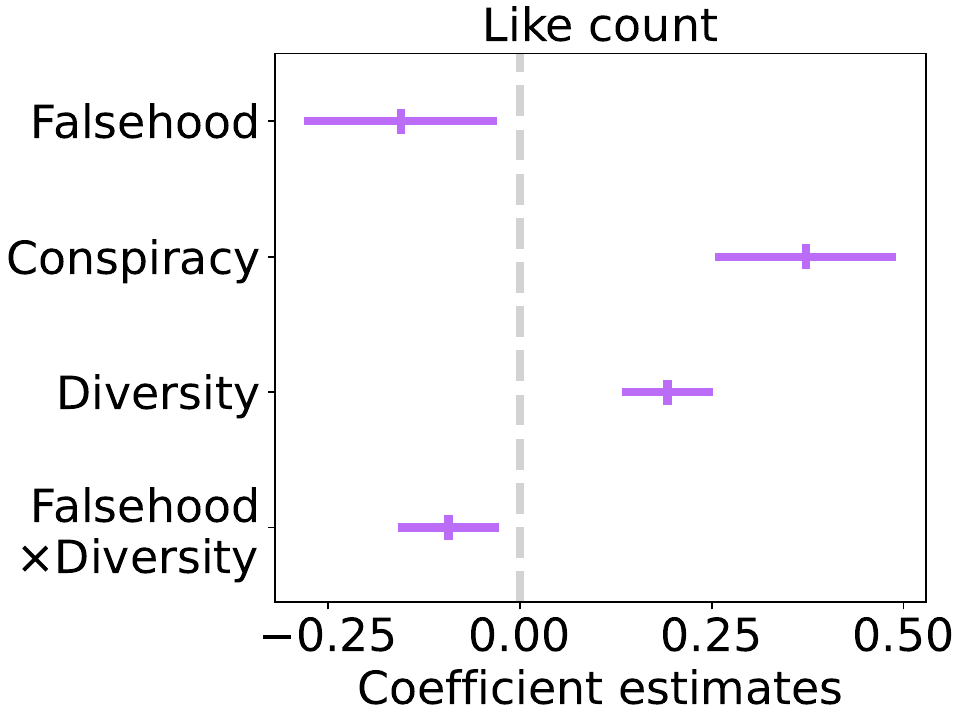}
\label{fig:like_count_coefs}
\end{subfigure}
\hfill
\begin{subfigure}{0.32\textwidth}
\caption{}
\includegraphics[width=\textwidth]{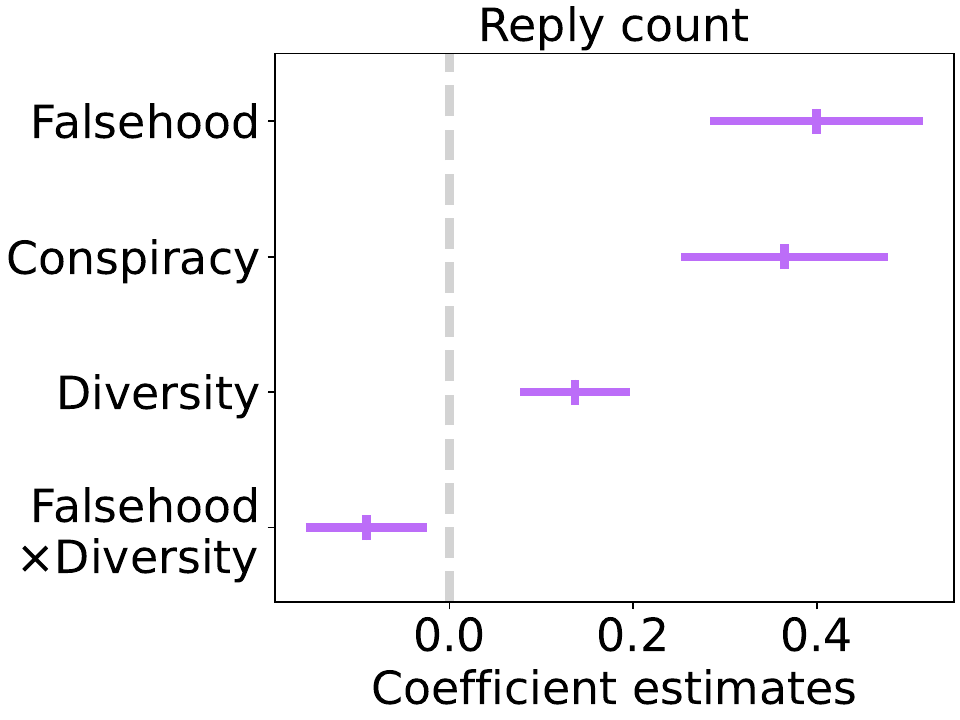}
\label{fig:reply_count_coefs}
\end{subfigure}
\caption{The engagement at post viewer level -- repost count, like count, and reply count. \subref{fig:ccdf_retweet_count}--\subref{fig:ccdf_reply_count} The CCDFs for \subref{fig:ccdf_retweet_count} repost count, \subref{fig:ccdf_like_count} like count, and \subref{fig:ccdf_reply_count} reply count. Three categories of CCDFs are included in each figure, \ie, true news (True), false news (False), and CT news (Consp). \subref{fig:repost_count_coefs}--\subref{fig:reply_count_coefs} The coefficient estimates of the independent variables -- $\var{Falsehood}$, $\var{Conspiracy}$, $\var{Diversity}$, and $\var{Falsehood} \times \var{Diversity}$. Other independent variables are included during estimation but omitted in the visualization for better readability. Shown are mean values with error bars representing 95\% CIs. The dependent variables are \subref{fig:repost_count_coefs} repost count, \subref{fig:like_count_coefs} like count, and \subref{fig:reply_count_coefs} reply count, respectively. See Table \ref{tab:post_engagement} in Suppl. \ref{sec:main_estimation} for full estimation results.}
\label{fig:post_engagement}
\end{figure}

\subsubsection{Descriptive statistics.}
In terms of engagement at post viewer level, we examine the numbers of reposts, likes, and replies for news posts. To this end, we first conduct descriptive analysis without incorporating the defined control variables from content and poster characteristics. We find that, on average, true news posts receive more engagement on X than false news posts in terms of reposts (Fig. \ref{fig:ccdf_retweet_count}; true: 29.396, false: 14.403, $\var{KS} =$ \num{0.240}, $p < $ \num{0.001}), likes (Fig. \ref{fig:ccdf_like_count}; true: 84.765, false: 45.234, $\var{KS} =$ \num{0.235}, $p < $ \num{0.001}), and replies (Fig. \ref{fig:ccdf_reply_count}; true: 6.294, false: 5.031, $\var{KS} =$ \num{0.133}, $p < $ \num{0.001}). However, it is worth noting that news posts that receive the most significant reposts, likes, or replies belong to the false category. Additionally, CT news posts receive more reposts (mean of 37.942, $\var{KS} =$ \num{0.043}, $p < $ \num{0.001}) and replies (mean of 25.173, $\var{KS} =$ \num{0.031}, $p < $ \num{0.01}) than overall false news posts, while without statistically significant difference in like count ($\var{KS} =$ \num{0.017}, $p = $ \num{0.194}).

\subsubsection{Regression results.}
Incorporating multiple control variables related to content and poster characteristics that can affect engagement with news posts, we conduct regression analysis to estimate the effects of $\var{Falsehood}$, $\var{Conspiracy}$, and $\var{Diversity}$ on the numbers of reposts, likes, and replies (Figs. \ref{fig:repost_count_coefs}--\ref{fig:reply_count_coefs}). In Fig. \ref{fig:repost_count_coefs}, the coefficient estimate of $\var{Falsehood}$ is significantly negative ($\var{coef.} =$ \num{-0.256}, $\var{p} <$ \num{0.001}), which means that false news posts receive fewer reposts than true news posts. On average, false news posts receive 22.6\% fewer reposts than true news posts. Similarly, the coefficient estimate of $\var{Falsehood}$ in Fig. \ref{fig:like_count_coefs} is significantly negative ($\var{coef.} =$ \num{-0.155}, $\var{p} <$ \num{0.05}). This means that, on average, false news posts receive 14.4\% fewer likes than true news posts. However, the coefficient estimate of $\var{Falsehood}$ in Fig. \ref{fig:reply_count_coefs} is significantly positive ($\var{coef.} =$ \num{0.400}, $\var{p} <$ \num{0.001}). This means that, on average, false news posts receive 49.2\% more replies than true news posts. 

Additionally, the coefficient estimates of $\var{Diversity}$ in Fig. \ref{fig:repost_count_coefs} ($\var{coef.} =$ \num{0.181}, $\var{p} <$ \num{0.001}), Fig. \ref{fig:like_count_coefs} ($\var{coef.} =$ \num{0.192}, $\var{p} <$ \num{0.001}), and Fig. \ref{fig:reply_count_coefs} ($\var{coef.} =$ \num{0.137}, $\var{p} <$ \num{0.001}) are significantly positive. This indicates that heightened topic diversity is associated with increased online engagement in reposts, likes, and replies. Specifically, a one standard deviation increase in topic diversity is associated with 19.8\% more reposts, 21.2\% more likes, and 14.7\% more replies. However, the coefficient estimates of $\var{Falsehood} \times \var{Diversity}$ in Fig. \ref{fig:repost_count_coefs} ($\var{coef.} =$ \num{-0.125}, $\var{p} <$ \num{0.001}), Fig. \ref{fig:like_count_coefs} ($\var{coef.} =$ \num{-0.093}, $\var{p} <$ \num{0.01}), and Fig. \ref{fig:reply_count_coefs} ($\var{coef.} =$ \num{-0.091}, $\var{p} <$ \num{0.01}) are significantly negative.\footnote{The coefficient estimates of $\var{Diversity}$ across repost count, like count, and reply count remain significantly positive in the subset analysis for false news posts. See Table \ref{tab:post_engagement_falsehood} in Suppl. \ref{sec:main_estimation} for full estimation results.} This suggests that the positive association of topic diversity with engagement on X is not significantly stronger within the context of false news posts compared to true news posts. On the contrary, the coefficient estimates of $\var{Conspiracy}$ in Fig. \ref{fig:repost_count_coefs} ($\var{coef.} =$ \num{0.342}, $\var{p} <$ \num{0.001}), Fig. \ref{fig:like_count_coefs} ($\var{coef.} =$ \num{0.373}, $\var{p} <$ \num{0.001}), and Fig. \ref{fig:reply_count_coefs} ($\var{coef.} =$ \num{0.365}, $\var{p} <$ \num{0.001}) are significantly positive. This means that the integration of conspiracy theories within false news posts is positively linked to engagement at post viewer level. Specifically, false news posts that contain conspiracy theories receive 40.8\% more reposts, 45.2\% more likes, and 44.1\% more replies compared to false news posts without conspiracy theories.

\textbf{Clarification for the discrepancies between descriptive statistics and regression results:}
We observe two discrepancies between descriptive analysis and regression analysis: (i) In the descriptive analysis, Fig. \ref{fig:ccdf_reply_count} shows that true news posts receive more replies than false news posts, whereas Fig. \ref{fig:reply_count_coefs} indicates that false news posts receive more replies than true news posts. (ii) No significant difference between CT news posts and overall false news posts is observed in the CCDF of like count (Fig. \ref{fig:ccdf_like_count}), while Fig. \ref{fig:like_count_coefs} indicates that false news posts containing conspiracy theories receive more likes than those without conspiracy theories. These discrepancies suggest that the falsehood may correlate with other variations in the control variables, potentially complicating the findings in the descriptive analysis and underscoring the necessity of incorporating control variables that can affect engagement.

For instance, emotions, included as control variables in our regression models, have been well documented as a significant driver in facilitating online post sharing \cite{vosoughi_spread_2018,chuai_anger_2022,chuai_what_2022,horner_emotions_2021,robertson_negativity_2023,solovev_moral_2022}. We find that falsehood is positively correlated with anger ($\var{r} =$ \num{0.038}, $\var{p} <$ \num{0.001}) and surprise ($\var{r} =$ \num{0.051}, $\var{p} <$ \num{0.001}) but negatively correlated with joy ($\var{r} =$ \num{-0.020}, $\var{p} <$ \num{0.001}; see Table \ref{tab:corrs_tweets} in Suppl. \ref{sec:corrs_vifs}). This suggests that false news posts express more anger and surprise but less joy compared to true news posts. Within the estimation results of our regression models, the coefficient estimates of $\var{Anger}$ for repost count ($\var{coef.} =$ \num{0.049}, $\var{p} <$ \num{0.001}), like count ($\var{coef.} =$ \num{0.059}, $\var{p} <$ \num{0.001}), and reply count ($\var{coef.} =$ \num{0.040}, $\var{p} <$ \num{0.01}) are consistently positive and statistically significant (see Table \ref{tab:post_engagement} in Suppl. \ref{sec:main_estimation} for full estimation results). Similarly, the coefficient estimates of $\var{Surprise}$ for repost count ($\var{coef.} =$ \num{0.093}, $\var{p} <$ \num{0.001}), like count ($\var{coef.} =$ \num{0.120}, $\var{p} <$ \num{0.001}), and reply count ($\var{coef.} =$ \num{0.121}, $\var{p} <$ \num{0.001}) are also consistently positive and statistically significant. Our results suggest that news posts expressing higher levels of anger or surprise tend to receive more reposts, likes, and replies, aligning with findings from previous research \cite{vosoughi_spread_2018,chuai_anger_2022}. However, the coefficient estimate of $\var{Joy}$ is not statistically significant for repost count, while significantly positive for reply count ($\var{coef.} =$ \num{0.077}, $\var{p} <$ \num{0.001}). This suggests that news posts expressing more joy are more likely to receive replies but do not receive more reposts. In light of this, we find two opposite effects: falsehood contributes to an increase in replies for false news posts, while the lower expression of joy in false news posts, compared to true news posts, links to fewer replies. The aggregation of these two opposite effects in the descriptive analysis could partially explain the discrepancy observed between Fig. \ref{fig:ccdf_reply_count} and Fig. \ref{fig:reply_count_coefs}.

In summary, at post viewer level, news posts with higher topic diversity receive more reposts, likes, and replies; however, this association is not stronger for false news posts (RQ 1.2). Additionally, false news posts containing conspiracy theories receive more reposts, likes, and replies compared to false news posts without conspiracy theories (RQ 2.2).

\subsection{Robustness Checks}
To ensure the robustness of our findings, we conduct several additional checks: (i) We examine the variance inflation factors (VIFs) among the independent variables at both news claim and news post levels (see Table \ref{tab:vifs} in Suppl. \ref{sec:corrs_vifs}). All the VIFs are significantly lower than the critical threshold of 4, which indicates that multicollinearity is not an issue for our analysis. (ii) To test whether our findings are robust with potential noise in topic keywords, we conduct our analysis again using the original 26 topics and associated keywords without manual validation (see details in Suppl. \ref{sec:original_topics}). (iii) To ensure that our findings are not influenced by the topic granularity, we recalculate topic diversity at the theme level and repeat our analysis (see details in Suppl. \ref{sec:theme_level}). (iv) We further balance the samples between true and false news items through propensity score matching and reduce potential unobserved confounding factors (see details in Suppl. \ref{sec:psm}). (v) We extend conspiracy theories by including QAnon and repeat our analysis (see details in Suppl. \ref{sec:qanon}).\footnote{QAnon is a far-right conspiracy theory alleging that a cabal of politicians and media personalities are engaged in a war to undermine society \cite{papasavva_gospel_2022,bar_finding_2023,kim_information_2023,moore_private_2024}. The QAnon discourse has become more diverse and evolved into a global phenomenon during the COVID-19 pandemic \cite{hoseini_globalization_2023}. We adopt a list of QAnon-specific keywords from previous research \cite{sharma_characterizing_2022} to identify QAnon conspiracy theory in our dataset. The performance of this keyword list in identifying QAnon discourse has been manually validated, showing high agreement with human assessments \cite{bar_finding_2023}. As a result, we identify 6 QAnon news claims and 302 QAnon news posts.} (vi) Given that news posts can also debunk linked false news items, we exclude potential debunking posts and repeat our analysis (see details in Suppl. \ref{sec:nodebunk}). (vii) We use an alternative topic modeling method (BERTopic) to cluster topics again and repeat our analysis (see details in Suppl. \ref{sec:bertopic}). All the results are robust and consistently support our main findings.

\subsection{Summary of Main Findings}
Our analysis yields several key findings: (i) False news, especially when accompanied by conspiracy theories, exhibits higher topic diversity compared to true news. This discrepancy increases from source news claims to news posts on X. (ii) At news source sharer level, neither the news lifetime nor the number of news posts is significantly associated with topic diversity. However, false news has a longer lifetime and receives more posts compared to true news. Additionally, the integration of conspiracy theories is associated with a longer lifetime for false news but has no significant association with the number of false news posts. (iii) At post viewer level, posts with higher topic diversity receive more reposts, likes, and replies, while this association is not stronger in false news posts. (iv) The integration of conspiracy theories is linked to more engagement with COVID-19 misinformation at post viewer level. False news posts that contain conspiracy theories, on average, receive 40.8\% more reposts, 45.2\% more likes, and 44.1\% more replies compared to false news posts without conspiracy theories.

\section{Discussion}
\subsection{Relevance}
The spread of misinformation surrounding health crises is a serious public concern, intersecting multiple domains in society. Here, we present an empirical analysis of \num{7273} fact-checked source news claims and their associated posts on X to explore how topic diversity and conspiracy theories influence engagement with multi-topic misinformation during the COVID-19 pandemic. Understanding the characteristics of misinformation and their effects on user engagement is a crucial and growing area of research for developing effective countermeasures and mitigating the spread of misinformation \cite{vosoughi_spread_2018,chuai_anger_2022,chuai_roll-out_2023,ecker_psychological_2022,ecker_misinformation_2024}. Our study makes two main contributions: (i) We analyze the characteristics of misinformation and its spread from the perspectives of topic diversity and conspiracy theories; (ii) We examine how these characteristics shape engagement with misinformation at two distinct user levels, \ie, news source sharer level and post viewer level. Specifically, we investigate online user engagement from the shares of source news items by news source sharers to the subsequent engagement with news posts by post viewers on X (\ie, reposts, likes, and replies). Our findings offer valuable insights for further research and practical strategies for combating the spread of online misinformation during health crises and beyond.

\subsection{Research Implications}

\subsubsection{Heightened topic diversity in misinformation/conspiracy theories.}
The observed high topic diversity in COVID-19 misinformation suggests that the interconnection of multiple topics becomes a characteristic of misinformation narratives during health crises. Here, the topic diversity reflects the interconnectedness of various topics within a single piece of misinformation, potentially enhancing its novelty. Different from our study, \citet{vosoughi_spread_2018} investigated the topic similarity between users' reposted true (or false) posts and their previously viewed posts. They found that false posts are more dissimilar from previously viewed posts compared to true posts. However, given users' topic interests and recommendation algorithms' tendency to promote similar content, users potentially consume more content that aligns with their interests \cite{cinus_effect_2022,engel_learning_2023}. These seemingly conflicting findings could be partially reconciled by our insights on topic diversity: misinformation may strategically link familiar topics together with diverse, novel topics, thereby drawing users' attention to new information and increasing their susceptibility. Particularly, our finding that misinformation containing conspiracy theories exhibits the strongest topic diversity underscores this deliberate strategy in conspiracy narratives. Moreover, previous research reveals coordinated efforts among misinformation websites from multiple domains to hyperlink to each other, further supporting our interpretation \cite{hanley_golden_2023}.

Additionally, we find that the topic diversity in COVID-19 misinformation significantly increases from its source claims to subsequent online posts. This mutation/evolution may contribute to the adaptability of misinformation, allowing it to resonate with diverse audiences and contexts \cite{yan_are_2022}. For instance, QAnon is a far-right political conspiracy theory that originated in 2017. It became more diverse during COVID-19, evolving and expanding beyond political themes to incorporate pandemic-related topics \cite{hoseini_globalization_2023}. Theoretically, within the context of multi-topic misinformation, individuals tend to adaptively connect related information and make associative inferences to reduce uncertainty and gain control in a way that is congruent to their prior knowledge \cite{lee_associative_2023,miani_interconnectedness_2022}. Understanding this cognitive process is essential for developing targeted interventions aimed at reducing susceptibility to multi-topic misinformation, especially when accompanied by conspiracy theories. Taken together, our findings underscore the importance of considering the multi-topic nature of misinformation in further research.

\subsubsection{Engagement patterns at news source sharer level and post viewer level.}
During health crises marked by high uncertainty, novel information can update users' understanding and knowledge of the current state, helping them manage anxiety and fostering their engagement intentions \cite{vosoughi_spread_2018,freiling_believing_2021,chuai_anger_2022,chen_what_2023}. Furthermore, sharing novel posts can make users feel that they possess rare and important information that others may lack, thereby boosting their self-esteem \cite{douglas_understanding_2019,chuai_anger_2022}. In this light, our findings suggest that the high topic diversity and the incorporation of conspiracy theories in misinformation can enhance its perceived novelty, thereby driving engagement. However, through further examining how topic diversity and conspiracy theories shape engagement at news source sharer level and post viewer level, we find distinct engagement patterns at the two user levels.

\textbf{Engagement patterns at news source sharer level:} At news source sharer level, the observed longer lifetime and higher post count on X for false news items, compared to true news items, indicate the persistence and widespread reach of misinformation on social media during COVID-19. The persistence of false news items provides them with more opportunities to accumulate engagement on social media, potentially reaching a broader audience. Moreover, the lifetime and post count appear to be largely attributed to the veracity of news claims, as opposed to other variables including $\var{Conspiracy}$ and $\var{Diversity}$. Given this, it is plausible that the prolonged lifetime and high post count for false news items mainly result from deliberately planned dissemination strategies by misinformation providers, involving means such as social bots and monetization \cite{teng_characterizing_2022,moffitt_hunting_2021,ballard_conspiracy_2022}. This perspective also aligns with the prior finding that social bots amplify low-credibility content during the initial spreading stages \cite{shao_spread_2018}. Consequently, the role of individuals' self-motivated sharing behaviors in the persistence and diffusion of COVID-19 misinformation is limited at news source sharer level. However, massive posts repeating the same false claims with adaptability -- \ie, through increased topic diversity -- can still heighten individuals' susceptibility and beliefs. In this regard, social media platforms should take measures to restrict repeated posts linking to the same false news items and their mutations.

\textbf{Engagement patterns at post viewer level:}
As opposed to engagement at news source sharer level, engagement at post viewer level (\ie, reposts, likes, and replies) on X might be largely driven by organic interactions from human users rather than social bots during COVID-19 \cite{teng_characterizing_2022}. Accordingly, we find that topic diversity and conspiracy theories become significant in driving engagement with misinformation at post viewer level in terms of reposts, likes, and replies. To efficiently reduce human users' engagement with misinformation at this level, social media platforms need to prioritize targeting misinformation that exhibits high topic diversity and incorporates conspiracy theories.

Additionally, our analysis reveals that replies differ from reposts and likes in their engagement with misinformation. Specifically, excluding the effects of topic diversity, conspiracy theories, and other control variables, false news posts receive more replies but fewer reposts and likes than true news posts. Distinct from reposts and likes, replies can not only support but also debunk false news posts. Given this, the higher number of replies for false news posts than true news posts can be triggered by debunking misinformation \cite{zannettou_won_2021}. Our supplementary analysis of quotes shows that, although not statistically significant, false news posts also have a trend to receive more quotes than true news posts (see Suppl. \ref{sec:quote_count} for full estimation results). This finding suggests that quotes serve a similar function as replies.

Taken together, our findings suggest that engagement with misinformation might be primarily driven by social bots at news source sharer level, while human users dominate engagement at post viewer level. This distinction highlights the need to examine misinformation engagement and develop intervention strategies at a multi-level framework.

\subsection{Practical Implications}
\subsubsection{Implementing cross-platform fact-checking.}
Currently, social media providers are actively implementing measures to control the spread of misinformation on their platforms, including third-party professional fact-checking by experts and community-based fact-checking driven by crowds \cite{chuai_roll-out_2023,ecker_psychological_2022}. However, these content moderation policies/features primarily function within individual platforms. Our findings suggest that misinformation from external sources, even if fact-checked on one platform, can still enter another platform through a large volume of posts. To prevent this spillover of misinformation across platforms, coordinated efforts among social media platforms, such as sharing fact-checking information, are needed.

\subsubsection{Expanding fact-checks to misinformation with same links.}
Given that misinformation can be repeatedly shared on social media platforms through the same source links, an effective strategy is to directly moderate posts that contain previous fact-checked misinformation links. For example, X has rolled out a Community Notes feature that allows users to write fact-checking notes as additional context displayed on the corresponding misleading posts.\footnote{More details are available at \url{https://communitynotes.x.com/guide/en/about/introduction}} This community-based fact-checking approach has proven effective in reducing the spread of misleading posts \cite{chuai_community_2024}. A recent update to Community Notes on X ensures that helpful notes that indicate posts' media elements, such as images or videos, are misleading will be visible on all posts containing the same media. Similarly, fact-checking notes about misinformation links could be expanded to cover all posts that contain the same links.

\subsubsection{Empowering logical reasoning and critical thinking for social media users.}
To reduce users' susceptibility to misinformation with high topic diversity and promote rational information processing, tools that can enhance users' logical reasoning and critical thinking abilities should be adopted. The rise of Large Language Models (LLMs) presents a promising opportunity to help users conduct logical reasoning and identify potentially conflicting or fictional information. Although current LLMs may still have limitations in reasoning tasks, particularly for misinformation identification \cite{pareek_effect_2024}, this approach represents a promising direction as LLMs continue to improve \cite{zavolokina_think_2024}.

\subsection{Limitations and Further Research} 
Our study is subject to several limitations. First, our analysis is based on real-world social media data and, as is typical with observational studies, is limited to reporting associations between variables rather than causal claims. Future research could benefit from experimental designs and user studies to establish causal relationships and explore heterogeneity in the effects of topic diversity and conspiracy theories on engagement across different user groups. 

Second, our findings on engagement can be influenced by social media recommendation algorithms and potential soft moderation implemented on misleading posts. Specifically, while our study mainly focuses on user-driven engagement at two levels and discusses the potential role of social bots, social media algorithms can selectively recommend posts and boost their visibility, potentially influencing user engagement and adding complexity to our results. Given that posts linking to the same news items can be treated similarly by recommendation algorithms or have similar audiences, the incorporation of news-specific random effects, \ie, random intercepts, in our models helps mitigate this concern. However, the precise role of recommendation algorithms warrants further exploration in future research.

Additionally, due to the absence of information indicating whether and when soft moderation was applied to certain misleading posts, we cannot control for the potential effect of soft moderation on subsequent engagement. Previous research has shown that soft moderation interventions -- \eg, attaching warning tags -- adopted by X during COVID-19 can reduce belief in misinformation and thus lower the spread/visibility of misleading posts \cite{sharevski_misinformation_2022}. Moreover, a recent study demonstrates that community-based fact-checking approach is effective in reducing the spread of misleading posts on X \cite{chuai_community_2024}. With the potential impact of such interventions in reducing engagement, we still find that false news has a longer lifetime than true news. This indicates that our results are not overestimated. On the other hand, prior research suggests that warning labels can also prompt interactions aimed at debunking false claims \cite{zannettou_won_2021}. To address this concern, we exclude potential debunking posts based on debunk-specific keywords and conduct our analysis again. Our results are robust and consistently support our main findings. 

Furthermore, the absence of detailed engagement timelines at post viewer level in our dataset hinders a more granular analysis of how topic diversity and conspiracy theories influence engagement with misinformation over time on social media. This limitation highlights an important area for further examination.

Finally, our analysis is based on a single curated public dataset that includes social context on one platform -- X \cite{li_mm-covid_2020}. Although this dataset offers unique features, such as multilingual content, it may introduce potential selection bias during data collection. To the best of our knowledge, no other publicly available datasets are suitable for addressing our research questions. Moreover, current restrictions on social media platform APIs make it increasingly challenging for researchers to access comprehensive data. We advocate for greater openness from social media platforms to facilitate academic research and encourage further studies to validate and generalize our findings across broader datasets and other social media platforms. 

\section{Conclusion}
The spread of health misinformation on social media, especially during health crises such as COVID-19, poses unprecedented threats to our society. It is crucial to understand the factors and mechanisms that contribute to the engagement with health misinformation on social media. Here, we provide a comprehensive analysis of how topic diversity and conspiracy theories shape engagement with COVID-19 misinformation from news source sharer level to post viewer level. We find that high topic diversity and the integration of conspiracy theories are characteristics of COVID-19 misinformation. Topic diversity has no significant association with the engagement at news source sharer level. However, topic diversity and conspiracy theories can significantly enhance engagement with COVID-19 misinformation at post viewer level. Our findings provide insights into understanding the engagement with health misinformation and highlight the engagement patterns for news source sharers and post viewers regarding topic diversity and conspiracy theories. These insights are valuable for developing targeted interventions at both user levels during health crises.

\section{Ethics Statement}
This research has received ethical approval from the Ethics Review Panel of the University of Luxembourg (ref. ERP 23-053 REMEDIS). 
All analyses are based on publicly available data. We declare no competing interests.

\begin{acks}
This research is supported by the Luxembourg National Research Fund (FNR) and Belgian National Fund for Scientific Research (FNRS), as part of the project REgulatory Solutions to MitigatE DISinformation (REMEDIS), grant ref. INTER\_FNRS\_21\_16554939\_REMEDIS.
\end{acks}

\bibliographystyle{ACM-Reference-Format}
\bibliography{refs}

\clearpage
\appendix
\begin{center}
    \Large \textbf{Supplementary Materials \\
    \small for ``From News Source Sharers to Post Viewers: How Topic Diversity and Conspiracy Theories Shape Engagement With Misinformation During a Health Crisis''}
\end{center}
\renewcommand\thetable{S\arabic{table}}
\setcounter{table}{0}
\renewcommand\thefigure{S\arabic{figure}}
\setcounter{figure}{0}
\renewcommand\thesection{S\arabic{section}}
\setcounter{section}{0}

\section{Estimation Results for Main Analysis}
\label{sec:main_estimation}

The full estimation results for the analysis in the main paper are reported in Tables \ref{tab:topic_diversity}--\ref{tab:post_engagement}. Specifically, Table \ref{tab:topic_diversity} reports the estimation results for the topic diversity. Table \ref{tab:news2tweets} reports the estimation results for the engagement at news source sharer level. Table \ref{tab:post_engagement} reports the estimation results for the engagement at post viewer level. Additionally, Table \ref{tab:post_engagement_falsehood} reports the estimation results for the subset analysis of post viewer engagement within false news posts.

\begin{table}[H]
\centering
\caption{Estimation results for the topic diversity in news claims [Column (1)], the topic diversity in news posts [Column (2)], and the changes in topic diversity from news claims to associated posts [Column (3)]. Reported are coefficient estimates with standard errors in parentheses. \sym{*} \(p<0.05\), \sym{**} \(p<0.01\), \sym{***} \(p<0.001\).}
\begin{tabularx}{\columnwidth}{@{\hspace{\tabcolsep}\extracolsep{\fill}}l*{3}{S}}
\toprule
&\multicolumn{1}{c}{(1)}&\multicolumn{1}{c}{(2)}&\multicolumn{1}{c}{(3)}\\
&\multicolumn{1}{c}{News claim}&\multicolumn{1}{c}{News post}&\multicolumn{1}{c}{Claim $\rightarrow$ Post}\\
\midrule
Falsehood   &       0.388\sym{***}&       0.751\sym{***}&       0.659\sym{***}\\
            &     (0.036)         &     (0.024)         &     (0.022)         \\
Conspiracy  &       0.061         &       0.136\sym{***}&       0.185\sym{***}\\
            &     (0.089)         &     (0.015)         &     (0.019)         \\
Words     &       0.391\sym{***}&       0.373\sym{***}&       0.390\sym{***}\\
            &     (0.013)         &     (0.004)         &     (0.004)         \\
Media       &                     &       0.022\sym{***}&       0.035\sym{***}\\
            &                     &     (0.006)         &     (0.008)         \\
Verified    &                     &       0.037\sym{***}&       0.004         \\
            &                     &     (0.008)         &     (0.010)         \\
AccountAge  &                     &       0.003         &       0.001         \\
            &                     &     (0.003)         &     (0.003)         \\
Followers   &                     &      -0.001         &      -0.019\sym{***}\\
            &                     &     (0.003)         &     (0.004)         \\
Followees   &                     &      -0.013\sym{***}&      -0.015\sym{***}\\
            &                     &     (0.002)         &     (0.003)         \\
Emotions       &       \checkmark&       \checkmark&       \checkmark\\
Language       &       \checkmark&       \checkmark&       \checkmark\\
MonthYear       &       \checkmark&       \checkmark&       \checkmark\\
Intercept  &      -0.078         &      -0.653\sym{***}&      -0.718\sym{***}\\
            &     (0.110)         &     (0.039)         &     (0.047)         \\
\midrule
News-level REs   &       {\xmark}         &          {\checkmark}           &     {\checkmark}                \\
\midrule
\(N\)       &        \num{7273}         &       \num{70904}         &       \num{70904}         \\
\(R^{2}\)   &       {0.179}         &          {\xmark}           &     {\xmark}                \\
\bottomrule
\end{tabularx}
\label{tab:topic_diversity}
\end{table}

\newpage
\begin{table}[H]
\centering
\caption{Estimation results for news lifetime [Column (1)] and post count [Column (2)]. Reported are coefficient estimates with standard errors in parentheses. \sym{*} \(p<0.05\), \sym{**} \(p<0.01\), \sym{***} \(p<0.001\).}
\begin{tabularx}{\columnwidth}{@{\hspace{\tabcolsep}\extracolsep{\fill}}l*{2}{S}}
\toprule
&\multicolumn{1}{c}{(1)}&\multicolumn{1}{c}{(2)}\\
&\multicolumn{1}{c}{News lifetime}&\multicolumn{1}{c}{Post count}\\
\midrule
Falsehood   &       1.239\sym{***}&       2.230\sym{***}\\
            &     (0.046)         &     (0.083)         \\
Conspiracy  &       0.452\sym{*}  &       0.493         \\
            &     (0.189)         &     (0.335)         \\
Diversity   &      -0.002         &      -0.025         \\
            &     (0.005)         &     (0.026)         \\
Falsehood $\times$ Diversity&      -0.075         &       0.079         \\
            &     (0.039)         &     (0.085)         \\
Words     &      -0.080\sym{***}&      -0.614\sym{***}\\
            &     (0.010)         &     (0.030)         \\
Emotions    &       \checkmark  &       \checkmark  \\    
Language    &       \checkmark  &       \checkmark  \\
MonthYear        &       \checkmark  &       \checkmark  \\
Intercept   &       2.491\sym{***}&       2.060\sym{***}\\
            &     (0.282)         &     (0.198)         \\
\midrule
\(N\)       &        \num{7273}         &       \num{7273}  \\
\(R^{2}\)   &       {0.481}         &          {0.188}           \\
\bottomrule
\end{tabularx}
\label{tab:news2tweets}
\end{table}

\begin{table}[H]
\centering
\caption{Estimation results for repost count [Column (1)], like count [Column (2)], and reply count [Column (3)]. Reported are coefficient estimates with standard errors in parentheses. \sym{*} \(p<0.05\), \sym{**} \(p<0.01\), \sym{***} \(p<0.001\).}
\begin{tabularx}{\columnwidth}{@{\hspace{\tabcolsep}\extracolsep{\fill}}l*{3}{S}}
\toprule
&\multicolumn{1}{c}{(1)}&\multicolumn{1}{c}{(2)}&\multicolumn{1}{c}{(3)}\\
&\multicolumn{1}{c}{Repost count}&\multicolumn{1}{c}{Like count}&\multicolumn{1}{c}{Reply count}\\
\midrule
Falsehood   &      -0.256\sym{***}&      -0.155\sym{*}  &       0.400\sym{***}\\
            &     (0.063)         &     (0.063)         &     (0.058)         \\
Conspiracy  &       0.342\sym{***}&       0.373\sym{***}&       0.365\sym{***}\\
            &     (0.065)         &     (0.059)         &     (0.056)         \\
Diversity&       0.181\sym{***}&       0.192\sym{***}&       0.137\sym{***}\\
            &     (0.031)         &     (0.030)         &     (0.030)         \\
Falsehood $\times$ Diversity&      -0.125\sym{***}&      -0.093\sym{**} &      -0.091\sym{**} \\
            &     (0.035)         &     (0.033)         &     (0.033)         \\
Anger       &       0.049\sym{***}&       0.059\sym{***}&       0.040\sym{**} \\
            &     (0.015)         &     (0.014)         &     (0.014)         \\
Disgust     &      -0.043\sym{**} &      -0.020         &      -0.020         \\
            &     (0.016)         &     (0.015)         &     (0.016)         \\
Fear        &       0.035\sym{*}  &       0.008         &       0.038\sym{*}  \\
            &     (0.016)         &     (0.015)         &     (0.015)         \\
Joy         &       0.006         &       0.030\sym{*}  &       0.077\sym{***}\\
            &     (0.016)         &     (0.015)         &     (0.015)         \\
Sadness     &       0.016         &      -0.009         &      -0.002         \\
            &     (0.017)         &     (0.015)         &     (0.016)         \\
Surprise    &       0.093\sym{***}&       0.120\sym{***}&       0.121\sym{***}\\
            &     (0.015)         &     (0.015)         &     (0.014)         \\
Words     &       0.519\sym{***}&       0.467\sym{***}&       0.443\sym{***}\\
            &     (0.017)         &     (0.016)         &     (0.015)         \\
Media       &       0.788\sym{***}&       0.596\sym{***}&       0.432\sym{***}\\
            &     (0.028)         &     (0.027)         &     (0.027)         \\
Verified    &       3.037\sym{***}&       3.071\sym{***}&       2.367\sym{***}\\
            &     (0.035)         &     (0.034)         &     (0.033)         \\
AccountAge  &       0.091\sym{***}&       0.084\sym{***}&       0.041\sym{***}\\
            &     (0.012)         &     (0.011)         &     (0.012)         \\
Followers   &       0.175\sym{***}&       0.180\sym{***}&       0.217\sym{***}\\
            &     (0.015)         &     (0.014)         &     (0.013)         \\
Followees   &       0.546\sym{***}&       0.383\sym{***}&       0.207\sym{***}\\
            &     (0.024)         &     (0.022)         &     (0.015)         \\
Language       &       \checkmark&       \checkmark&       \checkmark\\
MonthYear       &       \checkmark&       \checkmark&       \checkmark\\
Intercept  &       0.129         &       1.055\sym{***}&      -1.398\sym{***}\\
            &     (0.169)         &     (0.155)         &     (0.153)         \\
\midrule
News-level REs   &       {\checkmark}         &          {\checkmark}           &     {\checkmark} \\
\midrule
\(N\)       &        \num{70904}         &       \num{70904}         &       \num{70904}         \\
\(R^{2}\)   &       {\xmark}         &          {\xmark}           &       {\xmark}         \\
\bottomrule
\end{tabularx}
\label{tab:post_engagement}
\end{table}

\begin{table}[H]
\centering
\caption{Estimation results for repost count [Column (1)], like count [Column (2)], and reply count [Column (3)] in the subset of false news posts. Reported are coefficient estimates with standard errors in parentheses. \sym{*} \(p<0.05\), \sym{**} \(p<0.01\), \sym{***} \(p<0.001\).}
\begin{tabularx}{\columnwidth}{@{\hspace{\tabcolsep}\extracolsep{\fill}}l*{3}{S}}
\toprule
&\multicolumn{1}{c}{(1)}&\multicolumn{1}{c}{(2)}&\multicolumn{1}{c}{(3)}\\
&\multicolumn{1}{c}{Repost count}&\multicolumn{1}{c}{Like count}&\multicolumn{1}{c}{Reply count}\\
\midrule
Conspiracy  &       0.356\sym{***}&       0.389\sym{***}&       0.378\sym{***}\\
            &     (0.075)         &     (0.066)         &     (0.062)         \\
Diversity&       0.073\sym{***}&       0.134\sym{***}&       0.053\sym{**} \\
            &     (0.021)         &     (0.019)         &     (0.018)         \\
Words     &       0.502\sym{***}&       0.416\sym{***}&       0.454\sym{***}\\
            &     (0.025)         &     (0.022)         &     (0.021)         \\
Media       &       0.895\sym{***}&       0.712\sym{***}&       0.525\sym{***}\\
            &     (0.037)         &     (0.034)         &     (0.033)         \\
Verified    &       2.624\sym{***}&       2.564\sym{***}&       1.969\sym{***}\\
            &     (0.048)         &     (0.045)         &     (0.042)         \\
AccountAge  &       0.097\sym{***}&       0.113\sym{***}&       0.079\sym{***}\\
            &     (0.016)         &     (0.014)         &     (0.014)         \\
Followers   &       0.502\sym{***}&       0.616\sym{***}&       0.602\sym{***}\\
            &     (0.046)         &     (0.046)         &     (0.036)         \\
Followees   &       0.717\sym{***}&       0.512\sym{***}&       0.269\sym{***}\\
            &     (0.035)         &     (0.031)         &     (0.022)         \\
Emotions       &       \checkmark&       \checkmark&       \checkmark\\
Language       &       \checkmark&       \checkmark&       \checkmark\\
MonthYear       &       \checkmark&       \checkmark&       \checkmark\\
Intercept  &       0.129         &       1.290\sym{***}&      -0.451\sym{*}  \\
            &     (0.215)         &     (0.190)         &     (0.186)         \\
\midrule
News-level REs   &       {\checkmark}         &          {\checkmark}           &     {\checkmark} \\
\midrule
\(N\)       &        \num{52271}         &       \num{52271}         &       \num{52271}         \\
\(R^{2}\)   &       {\xmark}         &          {\xmark}           &       {\xmark}         \\
\bottomrule
\end{tabularx}
\label{tab:post_engagement_falsehood}
\end{table}

\newpage
\section{Cross-Correlations \& Variance Inflation Factors}
\label{sec:corrs_vifs}

We provide the cross-correlations to show how the independent variables relate to each other at the news claim and news post levels. Table \ref{tab:corrs_news} reports the cross-correlations among the variables in news claims. Table \ref{tab:corrs_tweets} reports the cross-correlations among the variables in news posts.

Additionally, we examine the Variance Inflation Factors (VIFs) among the independent variables to mitigate the concern about multicollinearity. The variance inflation factors are reported in Table \ref{tab:vifs}. We find that the variance inflation factors for the independent variables are all close to one and well below the critical threshold of four, which indicates that multicollinearity is not an issue of our analysis.

\begin{table}[H]
\centering
\caption{Correlations of variables in news claims}
\resizebox{\textwidth}{!}{
\begin{tabular}{l*{9}{S}}
\toprule
& {Falsehood} & {Conspiracy} & {Anger} & {Disgust} & {Fear} & {Joy} & {Sadness} & {Surprise} & {Words} \\
\midrule
{Falsehood} & 1.000 \\
{Conspiracy} & 0.221 & 1.000 \\
{Anger} & -0.010 & -0.018 & 1.000 \\
{Disgust} & 0.094 & 0.014 & 0.182 & 1.000\\
{Fear} & 0.030 & -0.018 & 0.159 & -0.027 & 1.000\\
{Joy} & -0.007 & 0.026 & -0.120 & -0.118 & -0.216 & 1.000 \\
{Sadness} & 0.082 & -0.005 & 0.094 & 0.108 & 0.283 & -0.185 & 1.000 \\
{Surprise} & -0.009 & -0.020 & 0.036 & 0.050 & -0.087 & -0.001 & 0.020 & 1.000 \\
{Words} & -0.371 & -0.063 & 0.020 & -0.047 & 0.077 & 0.100 & -0.016 & 0.017 & 1.000 \\
\bottomrule
\end{tabular}}
\label{tab:corrs_news}
\end{table}

\newpage
\begin{table}[H]
\caption{Correlations of variables in news posts}
\begin{subtable}{\textwidth}
\resizebox{\textwidth}{!}{
\begin{tabular}{l*{9}{S}}
\toprule
& {Falsehood} & {Conspiracy} & {Anger} & {Disgust} & {Fear} & {Joy} & {Sadness} & {Surprise} & {Words}\\
\midrule
{Falsehood} & 1.000\\
{Conspiracy} & 0.151 & 1.000 \\
{Anger} & 0.038 & 0.002 & 1.000\\
{Disgust} & 0.122 & -0.019 & 0.139 & 1.000\\
{Fear} & 0.075 & -0.054 & 0.089 & -0.063 & 1.000\\
{Joy} & -0.020 & 0.062 & -0.142 & -0.155 & -0.254 & 1.000\\
{Sadness} & 0.133 & -0.065 & 0.093 & 0.014 & 0.375 & -0.220 & 1.000\\
{Surprise} & 0.051 & -0.007 & 0.021 & -0.060 & -0.124 & 0.057 & -0.015 & 1.000 \\
{Words} & 0.155 & 0.016 & 0.060 & -0.017 & 0.046 & 0.033 & 0.055 & 0.057 & 1.000 \\
{Media} & -0.076 & -0.051 & -0.017 & -0.031 & 0.013 & 0.007 & -0.000 & -0.033 & 0.104 \\
{Verified} & -0.278 & -0.072 & -0.018 & -0.003 & -0.031 & 0.003 & -0.061 & -0.028 & 0.080 \\
{AccountAge} & -0.185 & -0.035 & -0.033 & -0.020 & -0.030 & -0.007 & -0.036 & -0.019 & -0.060 \\
{Followers} & -0.220 & -0.045 & -0.000 & -0.019 & -0.015 & 0.028 & -0.035 & -0.014 & 0.104 \\
{Followees} & -0.006 & -0.011 & -0.001 & 0.091 & -0.017 & -0.002 & -0.023 & -0.001 & -0.071 \\
\bottomrule
\end{tabular}}
\end{subtable}

\vspace{1em}

\begin{subtable}{.68\textwidth}
\resizebox{\textwidth}{!}{
\begin{tabular}{l*{9}{S}}
\toprule
{Continued} & {Media} & {Verified} & {AccountAge} & {Followers} & {Followees}\\
\midrule
{Falsehood} \\
{Conspiracy} \\
{Anger}  \\
{Disgust}\\
{Fear} \\
{Joy} \\
{Sadness} \\
{Surprise}\\
{Words} \\
{Media} & 1.000 \\
{Verified} & 0.218 & 1.000 \\
{AccountAge} & 0.047 & 0.321 & 1.000\\
{Followers} & 0.116 & 0.429 & 0.227 & 1.000\\
{Followees} & 0.006 & 0.053 & 0.105 & 0.044 & 1.000 \\
\bottomrule
\end{tabular}}
\end{subtable}
\label{tab:corrs_tweets}
\end{table}

\begin{table}[H]
\centering
\caption{VIFs}
\begin{tabular}{l*{2}{S}}
\toprule
&{News}&{Post}\\
\midrule
Falsehood&1.24&1.22\\
Conspiracy&1.05&1.04\\
Anger&1.07&1.05\\
Disgust&1.07&1.09\\
Fear&1.18&1.25\\
Joy&1.10&1.14\\
Sadness&1.13&1.21\\
Surprise&1.01&1.03\\
Words&1.19&1.09\\
Media&&1.06\\
Verified&&1.41\\
AccountAge&&1.16\\
Followers&&1.27\\
Followees&&1.03\\
\midrule
Mean VIF&1.12&1.15\\
\bottomrule
\end{tabular}
\label{tab:vifs}
\end{table}

\newpage
\section{Robustness Check With Original 26 Topics}
\label{sec:original_topics}

We adopt the original 26 topics and associated keywords directly from previous research \cite{chandrasekaran_topics_2020} and repeat the analysis. Table \ref{tab:topic_diversity_original} reports the estimation results for the topic diversity. Table \ref{tab:news2tweets_original} reports the estimation results for the engagement at news source sharer level. Table \ref{tab:post_engagement_original} reports the estimation results for the engagement at post viewer level. All the results are robust and consistently support our main findings.

\begin{table}[H]
\centering
\caption{Estimation results for the topic diversity in news claims [Column (1)], the topic diversity in news posts [Column (2)], and the changes in topic diversity from news claims to associated posts [Column (3)]. The topic diversity is recalculated based on the original 26 topics. Reported are coefficient estimates with standard errors in parentheses. \sym{*} \(p<0.05\), \sym{**} \(p<0.01\), \sym{***} \(p<0.001\).}
\begin{tabularx}{\columnwidth}{@{\hspace{\tabcolsep}\extracolsep{\fill}}l*{3}{S}}
\toprule
&\multicolumn{1}{c}{(1)}&\multicolumn{1}{c}{(2)}&\multicolumn{1}{c}{(3)}\\
&\multicolumn{1}{c}{News claim}&\multicolumn{1}{c}{News post}&\multicolumn{1}{c}{Claim $\rightarrow$ Post}\\
\midrule
Falsehood   &       0.404\sym{***}&       0.796\sym{***}&       0.724\sym{***}\\
            &     (0.034)         &     (0.023)         &     (0.022)         \\
Conspiracy  &       0.030         &       0.117\sym{***}&       0.165\sym{***}\\
            &     (0.095)         &     (0.015)         &     (0.019)         \\
Words     &       0.417\sym{***}&       0.393\sym{***}&       0.412\sym{***}\\
            &     (0.013)         &     (0.003)         &     (0.004)         \\
Media       &                     &       0.008         &       0.018\sym{*}  \\
            &                     &     (0.006)         &     (0.008)         \\
Verified    &                     &       0.043\sym{***}&       0.013         \\
            &                     &     (0.008)         &     (0.010)         \\
AccountAge  &                     &       0.002         &      -0.001         \\
            &                     &     (0.003)         &     (0.003)         \\
Followers   &                     &      -0.001         &      -0.021\sym{***}\\
            &                     &     (0.003)         &     (0.004)         \\
Followees   &                     &      -0.015\sym{***}&      -0.018\sym{***}\\
            &                     &     (0.002)         &     (0.003)         \\
Emotions       &       \checkmark&       \checkmark&       \checkmark\\
Language       &       \checkmark&       \checkmark&       \checkmark\\
MonthYear       &       \checkmark&       \checkmark&       \checkmark\\
Intercept  &      -0.078         &      -0.709\sym{***}&      -0.873\sym{***}\\
            &     (0.109)         &     (0.037)         &     (0.046)         \\
\midrule
News-level REs&{\xmark}&{\checkmark}&{\checkmark}\\
\midrule
\(N\)       &        \num{7479}         &       \num{73006}         &       \num{73006}         \\
\(R^{2}\)   &       {0.196}         &          {\xmark}           &       {\xmark}              \\
\bottomrule
\end{tabularx}
\label{tab:topic_diversity_original}
\end{table}

\newpage
\begin{table}[H]
\centering
\caption{Estimation results for news lifetime [Column (1)] and post count [Column (2)]. The topic diversity is recalculated based on the original 26 topics. Reported are coefficient estimates with standard errors in parentheses. \sym{*} \(p<0.05\), \sym{**} \(p<0.01\), \sym{***} \(p<0.001\).}
\begin{tabularx}{\columnwidth}{@{\hspace{\tabcolsep}\extracolsep{\fill}}l*{2}{S}}
\toprule
&\multicolumn{1}{c}{(1)}&\multicolumn{1}{c}{(2)}\\
&\multicolumn{1}{c}{News lifetime}&\multicolumn{1}{c}{Post count}\\
\midrule
Falsehood   &       1.232\sym{***}&       2.220\sym{***}\\
            &     (0.045)         &     (0.082)         \\
Conspiracy  &       0.412\sym{*}  &       0.486         \\
            &     (0.186)         &     (0.332)         \\
Diversity   &      -0.002         &      -0.045         \\
            &     (0.005)         &     (0.026)         \\
Falsehood $\times$ Diversity&      -0.069         &       0.084         \\
            &     (0.040)         &     (0.081)         \\
Words     &      -0.081\sym{***}&      -0.609\sym{***}\\
            &     (0.010)         &     (0.030)         \\
Emotions       &       \checkmark&       \checkmark\\
Language    &       \checkmark  &       \checkmark  \\
MonthYear        &       \checkmark  &       \checkmark  \\
Intercept   &       2.534\sym{***}&       2.062\sym{***}\\
            &     (0.274)         &     (0.197)         \\
\midrule
\(N\)       &        \num{7479}         &       \num{7479}  \\
\(R^{2}\)   &       {0.478}         &           {0.185}         \\
\bottomrule
\end{tabularx}
\label{tab:news2tweets_original}
\end{table}

\begin{table}[H]
\centering
\caption{Estimation results for repost count [Column (1)], like count [Column (2)], and reply count [Column (3)]. The topic diversity is recalculated based on the original 26 topics. Reported are coefficient estimates with standard errors in parentheses. \sym{*} \(p<0.05\), \sym{**} \(p<0.01\), \sym{***} \(p<0.001\).}
\begin{tabularx}{\columnwidth}{@{\hspace{\tabcolsep}\extracolsep{\fill}}l*{3}{S}}
\toprule
&\multicolumn{1}{c}{(1)}&\multicolumn{1}{c}{(2)}&\multicolumn{1}{c}{(3)}\\
&\multicolumn{1}{c}{Repost count}&\multicolumn{1}{c}{Like count}&\multicolumn{1}{c}{Reply count}\\
\midrule
Falsehood   &      -0.240\sym{***}&      -0.150\sym{*}  &       0.407\sym{***}\\
            &     (0.061)         &     (0.061)         &     (0.057)         \\
Conspiracy  &       0.335\sym{***}&       0.365\sym{***}&       0.361\sym{***}\\
            &     (0.064)         &     (0.059)         &     (0.056)         \\
Diversity&       0.160\sym{***}&       0.167\sym{***}&       0.118\sym{***}\\
            &     (0.030)         &     (0.029)         &     (0.029)         \\
Falsehood $\times$ Diversity&      -0.048         &      -0.034         &      -0.042         \\
            &     (0.034)         &     (0.032)         &     (0.033)         \\
Words     &       0.517\sym{***}&       0.467\sym{***}&       0.442\sym{***}\\
            &     (0.017)         &     (0.016)         &     (0.015)         \\
Media       &       0.774\sym{***}&       0.594\sym{***}&       0.426\sym{***}\\
            &     (0.027)         &     (0.026)         &     (0.026)         \\
Verified    &       3.027\sym{***}&       3.036\sym{***}&       2.368\sym{***}\\
            &     (0.035)         &     (0.034)         &     (0.032)         \\
AccountAge  &       0.094\sym{***}&       0.089\sym{***}&       0.044\sym{***}\\
            &     (0.012)         &     (0.011)         &     (0.011)         \\
Followers   &       0.186\sym{***}&       0.184\sym{***}&       0.221\sym{***}\\
            &     (0.015)         &     (0.014)         &     (0.013)         \\
Followees   &       0.464\sym{***}&       0.300\sym{***}&       0.189\sym{***}\\
            &     (0.022)         &     (0.019)         &     (0.014)         \\
Emotions       &       \checkmark&       \checkmark&       \checkmark\\
Language       &       \checkmark&       \checkmark&       \checkmark\\
MonthYear       &       \checkmark&       \checkmark&       \checkmark\\
Intercept  &       0.057         &       1.020\sym{***}&      -1.447\sym{***}\\
            &     (0.165)         &     (0.151)         &     (0.150)         \\
\midrule
News-level REs&{\checkmark}&{\checkmark}&{\checkmark}\\
\midrule
\(N\)       &        \num{73006}         &       \num{73006}         &       \num{73006}         \\
\(R^{2}\)   &       {\xmark}         &          {\xmark}           &       {\xmark}              \\
\bottomrule
\end{tabularx}
\label{tab:post_engagement_original}
\end{table}

\newpage
\section{Robustness Check at Theme Level}
\label{sec:theme_level}

We recalculate topic diversity at the theme level and repeat our analysis. Table \ref{tab:topic_diversity_theme} reports the estimation results for the topic diversity. Table \ref{tab:news2tweets_theme} reports the estimation results for the engagement at news source sharer level. Table \ref{tab:post_engagement_theme} reports the estimation results for the engagement at post viewer level. All the results are robust and consistently support our main findings.

\begin{table}[H]
\centering
\caption{Estimation results for the topic diversity in news claims [Column (1)], the topic diversity in news posts [Column (2)], and the changes in topic diversity from news claims to associated posts [Column (3)]. The topic diversity is recalculated at the theme level. Reported are coefficient estimates with standard errors in parentheses. \sym{*} \(p<0.05\), \sym{**} \(p<0.01\), \sym{***} \(p<0.001\).}
\begin{tabularx}{\columnwidth}{@{\hspace{\tabcolsep}\extracolsep{\fill}}l*{3}{S}}
\toprule
&\multicolumn{1}{c}{(1)}&\multicolumn{1}{c}{(2)}&\multicolumn{1}{c}{(3)}\\
&\multicolumn{1}{c}{News claim}&\multicolumn{1}{c}{News post}&\multicolumn{1}{c}{Claim $\rightarrow$ Post}\\
\midrule
Falsehood   &       0.311\sym{***}&       0.699\sym{***}&       0.616\sym{***}\\
            &     (0.036)         &     (0.026)         &     (0.023)         \\
Conspiracy  &       0.033         &       0.130\sym{***}&       0.173\sym{***}\\
            &     (0.092)         &     (0.016)         &     (0.019)         \\
Words     &       0.354\sym{***}&       0.337\sym{***}&       0.342\sym{***}\\
            &     (0.013)         &     (0.004)         &     (0.004)         \\
Media       &                     &       0.030\sym{***}&       0.040\sym{***}\\
            &                     &     (0.007)         &     (0.008)         \\
Verified    &                     &       0.033\sym{***}&      -0.002         \\
            &                     &     (0.008)         &     (0.010)         \\
AccountAge  &                     &       0.005\sym{*}  &       0.004         \\
            &                     &     (0.003)         &     (0.003)         \\
Followers   &                     &       0.002         &      -0.020\sym{***}\\
            &                     &     (0.003)         &     (0.004)         \\
Followees   &                     &      -0.014\sym{***}&      -0.015\sym{***}\\
            &                     &     (0.002)         &     (0.003)         \\
Emotions       &       \checkmark&       \checkmark&       \checkmark\\
Language       &       \checkmark&       \checkmark&       \checkmark\\
MonthYear       &       \checkmark&       \checkmark&       \checkmark\\
Intercept  &      -0.101         &      -0.517\sym{***}&      -0.600\sym{***}\\
            &     (0.110)         &     (0.040)         &     (0.047)         \\
\midrule
News-level REs&{\xmark}&{\checkmark}&{\checkmark}\\
\midrule
\(N\)       &        \num{7273}         &       \num{70904}         &       \num{70904}         \\
\(R^{2}\)   &       {0.147}         &          {\xmark}           &       {\xmark}              \\
\bottomrule
\end{tabularx}
\label{tab:topic_diversity_theme}
\end{table}

\newpage
\begin{table}[H]
\centering
\caption{Estimation results for news lifetime [Column (1)] and post count [Column (2)]. The topic diversity is recalculated at the theme level. Reported are coefficient estimates with standard errors in parentheses. \sym{*} \(p<0.05\), \sym{**} \(p<0.01\), \sym{***} \(p<0.001\).}
\begin{tabularx}{\columnwidth}{@{\hspace{\tabcolsep}\extracolsep{\fill}}l*{2}{S}}
\toprule
&\multicolumn{1}{c}{(1)}&\multicolumn{1}{c}{(2)}\\
&\multicolumn{1}{c}{News lifetime}&\multicolumn{1}{c}{Post count}\\
\midrule
Falsehood   &       1.235\sym{***}&       2.238\sym{***}\\
            &     (0.046)         &     (0.084)         \\
Conspiracy  &       0.448\sym{*}  &       0.485         \\
            &     (0.188)         &     (0.324)         \\
Diversity   &      -0.005         &      -0.048         \\
            &     (0.005)         &     (0.026)         \\
Falsehood $\times$ Diversity&      -0.043         &       0.044         \\
            &     (0.042)         &     (0.081)         \\
Words     &      -0.082\sym{***}&      -0.602\sym{***}\\
            &     (0.010)         &     (0.030)         \\
Emotions       &       \checkmark&       \checkmark\\
Language    &       \checkmark  &       \checkmark  \\
MonthYear        &       \checkmark  &       \checkmark  \\
Intercept   &       2.490\sym{***}&       2.076\sym{***}\\
            &     (0.282)         &     (0.199)         \\
\midrule
\(N\)       &        \num{7273}         &       \num{7273}  \\
\(R^{2}\)   &       {0.480}         &           {0.188}         \\
\bottomrule
\end{tabularx}
\label{tab:news2tweets_theme}
\end{table}

\begin{table}[H]
\centering
\caption{Estimation results for repost count [Column (1)], like count [Column (2)], and reply count [Column (3)]. The topic diversity is recalculated at the theme level. Reported are coefficient estimates with standard errors in parentheses. \sym{*} \(p<0.05\), \sym{**} \(p<0.01\), \sym{***} \(p<0.001\).}
\begin{tabularx}{\columnwidth}{@{\hspace{\tabcolsep}\extracolsep{\fill}}l*{3}{S}}
\toprule
&\multicolumn{1}{c}{(1)}&\multicolumn{1}{c}{(2)}&\multicolumn{1}{c}{(3)}\\
&\multicolumn{1}{c}{Repost count}&\multicolumn{1}{c}{Like count}&\multicolumn{1}{c}{Reply count}\\
\midrule
Falsehood   &      -0.225\sym{***}&      -0.125\sym{*}  &       0.415\sym{***}\\
            &     (0.062)         &     (0.062)         &     (0.057)         \\
Conspiracy  &       0.346\sym{***}&       0.378\sym{***}&       0.368\sym{***}\\
            &     (0.065)         &     (0.059)         &     (0.056)         \\
Diversity&       0.159\sym{***}&       0.175\sym{***}&       0.125\sym{***}\\
            &     (0.030)         &     (0.028)         &     (0.028)         \\
Falsehood $\times$ Diversity&      -0.124\sym{***}&      -0.094\sym{**} &      -0.074\sym{*}  \\
            &     (0.034)         &     (0.032)         &     (0.031)         \\
Words     &       0.528\sym{***}&       0.477\sym{***}&       0.444\sym{***}\\
            &     (0.017)         &     (0.016)         &     (0.015)         \\
Media       &       0.786\sym{***}&       0.596\sym{***}&       0.431\sym{***}\\
            &     (0.028)         &     (0.027)         &     (0.027)         \\
Verified    &       3.036\sym{***}&       3.070\sym{***}&       2.366\sym{***}\\
            &     (0.035)         &     (0.034)         &     (0.033)         \\
AccountAge  &       0.091\sym{***}&       0.084\sym{***}&       0.041\sym{***}\\
            &     (0.012)         &     (0.011)         &     (0.012)         \\
Followers   &       0.175\sym{***}&       0.180\sym{***}&       0.217\sym{***}\\
            &     (0.015)         &     (0.014)         &     (0.013)         \\
Followees   &       0.546\sym{***}&       0.383\sym{***}&       0.207\sym{***}\\
            &     (0.024)         &     (0.022)         &     (0.015)         \\
Emotions       &       \checkmark&       \checkmark&       \checkmark\\
Language       &       \checkmark&       \checkmark&       \checkmark\\
MonthYear       &       \checkmark&       \checkmark&       \checkmark\\
Intercept  &       0.098         &       1.023\sym{***}&      -1.418\sym{***}\\
            &     (0.169)         &     (0.154)         &     (0.153)         \\
\midrule
News-level REs&{\checkmark}&{\checkmark}&{\checkmark}\\
\midrule
\(N\)       &        \num{70904}         &       \num{70904}         &       \num{70904}         \\
\(R^{2}\)   &       {\xmark}         &          {\xmark}           &       {\xmark}              \\
\bottomrule
\end{tabularx}
\label{tab:post_engagement_theme}
\end{table}

\newpage
\section{Robustness Check With Propensity Score Matching}
\label{sec:psm}

To reduce potential unobserved confounding factors and ensure the robustness of our results, we further conduct propensity score matching for true and false news claims. Specifically, we match true and false news claims based on the variables of emotions and the number of words, with settings including the caliper of 0.3, ``common support'', and ``noreplacement.'' Consequently, we discard 30 false news claims and achieve all biases lower than 0.05. Subsequently, we repeat our analysis based on the matched true and false news claims and their associated posts on X. Table \ref{tab:topic_diversity_psm} reports the estimation results for the topic diversity. Table \ref{tab:news2tweets_psm} reports the estimation results for the engagement at news source sharer level. Table \ref{tab:post_engagement_psm} reports the estimation results for the engagement at post viewer level. All the results are robust and consistently support our main findings.

\begin{table}[H]
\centering
\caption{Estimation results for the topic diversity in news claims [Column (1)], the topic diversity in news posts [Column (2)], and the changes in topic diversity from news claims to associated posts [Column (3)]. The estimations are based on samples obtained after propensity score matching. Reported are coefficient estimates with standard errors in parentheses. \sym{*} \(p<0.05\), \sym{**} \(p<0.01\), \sym{***} \(p<0.001\).}
\begin{tabularx}{\columnwidth}{@{\hspace{\tabcolsep}\extracolsep{\fill}}l*{3}{S}}
\toprule
&\multicolumn{1}{c}{(1)}&\multicolumn{1}{c}{(2)}&\multicolumn{1}{c}{(3)}\\
&\multicolumn{1}{c}{News claim}&\multicolumn{1}{c}{News post}&\multicolumn{1}{c}{Claim $\rightarrow$ Post}\\
\midrule
Falsehood   &       0.400\sym{***}&       0.745\sym{***}&       0.640\sym{***}\\
            &     (0.036)         &     (0.024)         &     (0.022)         \\
Conspiracy  &       0.037         &       0.137\sym{***}&       0.186\sym{***}\\
            &     (0.091)         &     (0.015)         &     (0.019)         \\
Words     &       0.387\sym{***}&       0.369\sym{***}&       0.383\sym{***}\\
            &     (0.013)         &     (0.004)         &     (0.004)         \\
Media       &                     &       0.021\sym{***}&       0.034\sym{***}\\
            &                     &     (0.006)         &     (0.008)         \\
Verified    &                     &       0.041\sym{***}&       0.009         \\
            &                     &     (0.008)         &     (0.010)         \\
AccountAge  &                     &       0.003         &       0.001         \\
            &                     &     (0.003)         &     (0.003)         \\
Followers   &                     &      -0.001         &      -0.019\sym{***}\\
            &                     &     (0.003)         &     (0.004)         \\
Followees   &                     &      -0.010\sym{***}&      -0.011\sym{***}\\
            &                     &     (0.002)         &     (0.003)         \\
Emotions       &       \checkmark&       \checkmark&       \checkmark\\
Language       &       \checkmark&       \checkmark&       \checkmark\\
MonthYear       &       \checkmark&       \checkmark&       \checkmark\\
Intercept  &      -0.075         &      -0.644\sym{***}&      -0.705\sym{***}\\
            &     (0.111)         &     (0.039)         &     (0.047)         \\
\midrule
News-level REs&{\xmark}&{\checkmark}&{\checkmark}\\
\midrule
\(N\)       &        \num{7243}         &       \num{69366}         &       \num{69366}         \\
\(R^{2}\)   &       {0.181}         &         {\xmark}            &       {\xmark}              \\
\bottomrule
\end{tabularx}
\label{tab:topic_diversity_psm}
\end{table}

\newpage
\begin{table}[H]
\centering
\caption{Estimation results for news lifetime [Column (1)] and post count [Column (2)]. The estimations are based on samples obtained after propensity score matching. Reported are coefficient estimates with standard errors in parentheses. \sym{*} \(p<0.05\), \sym{**} \(p<0.01\), \sym{***} \(p<0.001\).}
\begin{tabularx}{\columnwidth}{@{\hspace{\tabcolsep}\extracolsep{\fill}}l*{2}{S}}
\toprule
&\multicolumn{1}{c}{(1)}&\multicolumn{1}{c}{(2)}\\
&\multicolumn{1}{c}{News lifetime}&\multicolumn{1}{c}{Post count}\\
\midrule
Falsehood   &       1.234\sym{***}&       2.230\sym{***}\\
            &     (0.046)         &     (0.083)         \\
Conspiracy  &       0.479\sym{*}  &       0.501         \\
            &     (0.189)         &     (0.337)         \\
Diversity   &      -0.000         &      -0.024         \\
            &     (0.005)         &     (0.026)         \\
Falsehood $\times$ Diversity&      -0.073         &       0.077         \\
            &     (0.040)         &     (0.086)         \\
Words     &      -0.082\sym{***}&      -0.615\sym{***}\\
            &     (0.010)         &     (0.030)         \\
Emotions       &       \checkmark&       \checkmark\\
Language    &       \checkmark  &       \checkmark  \\
MonthYear        &       \checkmark  &       \checkmark  \\
Intercept   &       2.572\sym{***}&       2.077\sym{***}\\
            &     (0.275)         &     (0.197)         \\
\midrule
\(N\)       &        \num{7243}         &        \num{7243}         \\
\(R^{2}\)   &       {0.482}         &         {0.188}            \\
\bottomrule
\end{tabularx}
\label{tab:news2tweets_psm}
\end{table}

\begin{table}[H]
\centering
\caption{Estimation results for repost count [Column (1)], like count [Column (2)], and reply count [Column (3)]. The estimations are based on samples obtained after propensity score matching. Reported are coefficient estimates with standard errors in parentheses. \sym{*} \(p<0.05\), \sym{**} \(p<0.01\), \sym{***} \(p<0.001\).}
\begin{tabularx}{\columnwidth}{@{\hspace{\tabcolsep}\extracolsep{\fill}}l*{3}{S}}
\toprule
&\multicolumn{1}{c}{(1)}&\multicolumn{1}{c}{(2)}&\multicolumn{1}{c}{(3)}\\
&\multicolumn{1}{c}{Repost count}&\multicolumn{1}{c}{Like count}&\multicolumn{1}{c}{Reply count}\\
\midrule
Falsehood   &      -0.244\sym{***}&      -0.134\sym{*}  &       0.411\sym{***}\\
            &     (0.063)         &     (0.063)         &     (0.058)         \\
Conspiracy  &       0.360\sym{***}&       0.391\sym{***}&       0.370\sym{***}\\
            &     (0.065)         &     (0.059)         &     (0.057)         \\
Diversity&       0.183\sym{***}&       0.193\sym{***}&       0.133\sym{***}\\
            &     (0.031)         &     (0.030)         &     (0.030)         \\
Falsehood $\times$ Diversity&      -0.125\sym{***}&      -0.088\sym{**} &      -0.088\sym{**} \\
            &     (0.035)         &     (0.033)         &     (0.033)         \\
Words     &       0.512\sym{***}&       0.459\sym{***}&       0.442\sym{***}\\
            &     (0.017)         &     (0.016)         &     (0.015)         \\
Media       &       0.788\sym{***}&       0.597\sym{***}&       0.432\sym{***}\\
            &     (0.028)         &     (0.027)         &     (0.027)         \\
Verified    &       3.075\sym{***}&       3.118\sym{***}&       2.405\sym{***}\\
            &     (0.036)         &     (0.035)         &     (0.033)         \\
AccountAge  &       0.088\sym{***}&       0.081\sym{***}&       0.038\sym{**} \\
            &     (0.012)         &     (0.011)         &     (0.012)         \\
Followers   &       0.172\sym{***}&       0.177\sym{***}&       0.216\sym{***}\\
            &     (0.015)         &     (0.014)         &     (0.013)         \\
Followees   &       0.635\sym{***}&       0.467\sym{***}&       0.250\sym{***}\\
            &     (0.026)         &     (0.024)         &     (0.017)         \\
Emotions       &       \checkmark&       \checkmark&       \checkmark\\
Language       &       \checkmark&       \checkmark&       \checkmark\\
MonthYear       &       \checkmark&       \checkmark&       \checkmark\\
Intercept  &       0.061         &       0.974\sym{***}&      -1.475\sym{***}\\
            &     (0.169)         &     (0.155)         &     (0.154)         \\
\midrule
News-level REs&{\checkmark}&{\checkmark}&{\checkmark}\\
\midrule
\(N\)       &       \num{69366}         &       \num{69366}         &       \num{69366}         \\
\(R^{2}\)   &       {\xmark}         &          {\xmark}           &       {\xmark}              \\
\bottomrule
\end{tabularx}
\label{tab:post_engagement_psm}
\end{table}

\newpage
\section{Robustness Check With Inclusion of QAnon Conspiracy Theory}
\label{sec:qanon}

We further include QAnon to extend conspiracy theories and repeat our analysis. QAnon is a far-right conspiracy theory alleging that a cabal of politicians and media personalities are engaged in a war to undermine society \cite{papasavva_gospel_2022,bar_finding_2023,kim_information_2023}. The QAnon discourse has become more diverse and evolved into a global phenomenon during the COVID-19 pandemic \cite{hoseini_globalization_2023}. We adopt a list of QAnon-specific keywords from previous research \cite{sharma_characterizing_2022} to identify QAnon conspiracy theory in our dataset. The performance of this keyword list in identifying QAnon discourse has been manually validated, showing high agreement with human assessments \cite{bar_finding_2023}. As a result, we identify 6 QAnon news claims and 302 QAnon news posts. Table \ref{tab:topic_diversity_qanon} reports the estimation results for the topic diversity. Table \ref{tab:news2tweets_qanon} reports the estimation results for the engagement at news source sharer level. Table \ref{tab:post_engagement_qanon} reports the estimation results for the engagement at post viewer level. All the results are robust and consistently support our main findings.

\begin{table}[H]
\centering
\caption{Estimation results for the topic diversity in news claims [Column (1)], the topic diversity in news posts [Column (2)], and the changes in topic diversity from news claims to associated posts [Column (3)]. QAnon conspiracy theory is included during estimation. Reported are coefficient estimates with standard errors in parentheses. \sym{*} \(p<0.05\), \sym{**} \(p<0.01\), \sym{***} \(p<0.001\).}
\begin{tabularx}{\columnwidth}{@{\hspace{\tabcolsep}\extracolsep{\fill}}l*{3}{S}}
\toprule
&\multicolumn{1}{c}{(1)}&\multicolumn{1}{c}{(2)}&\multicolumn{1}{c}{(3)}\\
&\multicolumn{1}{c}{News claim}&\multicolumn{1}{c}{News post}&\multicolumn{1}{c}{Claim $\rightarrow$ Post}\\
\midrule
Falsehood   &       0.389\sym{***}&       0.751\sym{***}&       0.659\sym{***}\\
            &     (0.036)         &     (0.024)         &     (0.022)         \\
Conspiracy  &       0.030         &       0.124\sym{***}&       0.175\sym{***}\\
            &     (0.087)         &     (0.015)         &     (0.018)         \\
Words     &       0.391\sym{***}&       0.373\sym{***}&       0.389\sym{***}\\
            &     (0.013)         &     (0.004)         &     (0.004)         \\
Media       &                     &       0.022\sym{***}&       0.034\sym{***}\\
            &                     &     (0.006)         &     (0.008)         \\
Verified    &                     &       0.037\sym{***}&       0.005         \\
            &                     &     (0.008)         &     (0.010)         \\
AccountAge  &                     &       0.003         &       0.001         \\
            &                     &     (0.003)         &     (0.003)         \\
Followers   &                     &      -0.001         &      -0.019\sym{***}\\
            &                     &     (0.003)         &     (0.004)         \\
Followees   &                     &      -0.013\sym{***}&      -0.015\sym{***}\\
            &                     &     (0.002)         &     (0.003)         \\
Emotions       &       \checkmark&       \checkmark&       \checkmark\\
Language       &       \checkmark&       \checkmark&       \checkmark\\
MonthYear       &       \checkmark&       \checkmark&       \checkmark\\
Intercept  &      -0.075         &      -0.652\sym{***}&      -0.717\sym{***}\\
            &     (0.110)         &     (0.039)         &     (0.047)         \\
\midrule
News-level REs&{\xmark}&{\checkmark}&{\checkmark}\\
\midrule
\(N\)       &        \num{7273}         &       \num{70904}         &       \num{70904}         \\
\(R^{2}\)   &       {0.179}         &          {\xmark}           &       {\xmark}              \\
\bottomrule
\end{tabularx}
\label{tab:topic_diversity_qanon}
\end{table}

\newpage
\begin{table}[H]
\centering
\caption{Estimation results for news lifetime [Column (1)] and post count [Column (2)]. QAnon conspiracy theory is included during estimation. Reported are coefficient estimates with standard errors in parentheses. \sym{*} \(p<0.05\), \sym{**} \(p<0.01\), \sym{***} \(p<0.001\).}
\begin{tabularx}{\columnwidth}{@{\hspace{\tabcolsep}\extracolsep{\fill}}l*{2}{S}}
\toprule
&\multicolumn{1}{c}{(1)}&\multicolumn{1}{c}{(2)}\\
&\multicolumn{1}{c}{News lifetime}&\multicolumn{1}{c}{Post count}\\
\midrule
Falsehood   &       1.234\sym{***}&       2.232\sym{***}\\
            &     (0.046)         &     (0.083)         \\
Conspiracy  &       0.505\sym{**} &       0.443         \\
            &     (0.183)         &     (0.330)         \\
Diversity   &      -0.002         &      -0.025         \\
            &     (0.005)         &     (0.026)         \\
Falsehood $\times$ Diversity&      -0.074         &       0.080         \\
            &     (0.039)         &     (0.085)         \\
Words     &      -0.080\sym{***}&      -0.613\sym{***}\\
            &     (0.010)         &     (0.030)         \\
Emotions       &       \checkmark&       \checkmark\\
Language    &       \checkmark  &       \checkmark  \\
MonthYear        &       \checkmark  &       \checkmark  \\
Intercept   &       2.487\sym{***}&       2.065\sym{***}\\
            &     (0.282)         &     (0.198)         \\
\midrule
\(N\)       &        \num{7273}         &       \num{7273}  \\
\(R^{2}\)   &       {0.481}         &           {0.188}         \\
\bottomrule
\end{tabularx}
\label{tab:news2tweets_qanon}
\end{table}

\begin{table}[H]
\centering
\caption{Estimation results for repost count [Column (1)], like count [Column (2)], and reply count [Column (3)]. QAnon conspiracy theory is included during estimation. Reported are coefficient estimates with standard errors in parentheses. \sym{*} \(p<0.05\), \sym{**} \(p<0.01\), \sym{***} \(p<0.001\).}
\begin{tabularx}{\columnwidth}{@{\hspace{\tabcolsep}\extracolsep{\fill}}l*{3}{S}}
\toprule
&\multicolumn{1}{c}{(1)}&\multicolumn{1}{c}{(2)}&\multicolumn{1}{c}{(3)}\\
&\multicolumn{1}{c}{Repost count}&\multicolumn{1}{c}{Like count}&\multicolumn{1}{c}{Reply count}\\
\midrule
Falsehood   &      -0.251\sym{***}&      -0.150\sym{*}  &       0.403\sym{***}\\
            &     (0.063)         &     (0.063)         &     (0.058)         \\
Conspiracy  &       0.271\sym{***}&       0.300\sym{***}&       0.312\sym{***}\\
            &     (0.063)         &     (0.058)         &     (0.055)         \\
Diversity&       0.181\sym{***}&       0.192\sym{***}&       0.137\sym{***}\\
            &     (0.031)         &     (0.030)         &     (0.030)         \\
Falsehood $\times$ Diversity&      -0.125\sym{***}&      -0.092\sym{**} &      -0.090\sym{**} \\
            &     (0.035)         &     (0.033)         &     (0.033)         \\
Words     &       0.520\sym{***}&       0.468\sym{***}&       0.443\sym{***}\\
            &     (0.017)         &     (0.016)         &     (0.015)         \\
Media       &       0.788\sym{***}&       0.596\sym{***}&       0.432\sym{***}\\
            &     (0.028)         &     (0.027)         &     (0.027)         \\
Verified    &       3.037\sym{***}&       3.071\sym{***}&       2.367\sym{***}\\
            &     (0.035)         &     (0.034)         &     (0.033)         \\
AccountAge  &       0.092\sym{***}&       0.085\sym{***}&       0.041\sym{***}\\
            &     (0.012)         &     (0.011)         &     (0.012)         \\
Followers   &       0.175\sym{***}&       0.179\sym{***}&       0.217\sym{***}\\
            &     (0.015)         &     (0.014)         &     (0.013)         \\
Followees   &       0.545\sym{***}&       0.383\sym{***}&       0.207\sym{***}\\
            &     (0.024)         &     (0.022)         &     (0.015)         \\
Emotions       &       \checkmark&       \checkmark&       \checkmark\\
Language       &       \checkmark&       \checkmark&       \checkmark\\
MonthYear       &       \checkmark&       \checkmark&       \checkmark\\
Intercept  &       0.130         &       1.055\sym{***}&      -1.396\sym{***}\\
            &     (0.169)         &     (0.155)         &     (0.153)         \\
\midrule
News-level REs   &       {\checkmark}         &          {\checkmark}           &     {\checkmark} \\
\midrule
\(N\)       &        \num{70904}         &       \num{70904}         &       \num{70904}         \\
\(R^{2}\)   &       {\xmark}         &          {\xmark}           &       {\xmark}         \\
\bottomrule
\end{tabularx}
\label{tab:post_engagement_qanon}
\end{table}

\newpage
\section{Robustness Check With Exclusion of Potential Debunking Posts}
\label{sec:nodebunk}

News posts can also debunk the linked news items. We use keywords of ``misinformation,'' ``misleading,'' ``fake,'' ``false,'' and ``disinformation'' to identify potential debunking posts, resulting in \num{4316} news posts. We exclude these potential debunking posts and repeat our analysis. Table \ref{tab:topic_diversity_nodebunk} reports the estimation results for the topic diversity. Table \ref{tab:news2tweets_nodebunk} reports the estimation results for the engagement at news source sharer level. Table \ref{tab:post_engagement_nodebunk} reports the estimation results for the engagement at post viewer level. All the results are robust and consistently support our main findings.

\begin{table}[H]
\centering
\caption{Estimation results for the topic diversity in news claims [Column (1)], the topic diversity in news posts [Column (2)], and the changes in topic diversity from news claims to associated posts [Column (3)]. The potential debunking posts are excluded during estimation. Reported are coefficient estimates with standard errors in parentheses. \sym{*} \(p<0.05\), \sym{**} \(p<0.01\), \sym{***} \(p<0.001\).}
\begin{tabularx}{\columnwidth}{@{\hspace{\tabcolsep}\extracolsep{\fill}}l*{3}{S}}
\toprule
&\multicolumn{1}{c}{(1)}&\multicolumn{1}{c}{(2)}&\multicolumn{1}{c}{(3)}\\
&\multicolumn{1}{c}{News claim}&\multicolumn{1}{c}{News post}&\multicolumn{1}{c}{Claim $\rightarrow$ Post}\\
\midrule
Falsehood   &       0.364\sym{***}&       0.773\sym{***}&       0.717\sym{***}\\
            &     (0.037)         &     (0.025)         &     (0.023)         \\
Conspiracy  &       0.065         &       0.124\sym{***}&       0.171\sym{***}\\
            &     (0.095)         &     (0.016)         &     (0.020)         \\
Words     &       0.381\sym{***}&       0.370\sym{***}&       0.384\sym{***}\\
            &     (0.013)         &     (0.004)         &     (0.004)         \\
Media       &                     &       0.021\sym{**} &       0.033\sym{***}\\
            &                     &     (0.007)         &     (0.008)         \\
Verified    &                     &       0.049\sym{***}&       0.018         \\
            &                     &     (0.009)         &     (0.011)         \\
AccountAge  &                     &       0.002         &       0.001         \\
            &                     &     (0.003)         &     (0.003)         \\
Followers   &                     &      -0.005         &      -0.026\sym{***}\\
            &                     &     (0.003)         &     (0.004)         \\
Followees   &                     &      -0.009\sym{***}&      -0.010\sym{**} \\
            &                     &     (0.002)         &     (0.003)         \\
Emotions       &       \checkmark&       \checkmark&       \checkmark\\
Language       &       \checkmark&       \checkmark&       \checkmark\\
MonthYear       &       \checkmark&       \checkmark&       \checkmark\\
Intercept  &      -0.104         &      -0.636\sym{***}&      -0.697\sym{***}\\
            &     (0.117)         &     (0.039)         &     (0.048)         \\
\midrule
News-level REs&{\xmark}&{\checkmark}&{\checkmark}\\
\midrule
\(N\)       &        \num{7097}         &       \num{66588}         &       \num{66588}         \\
\(R^{2}\)   &       {0.173}         &          {\xmark}           &       {\xmark}              \\
\bottomrule
\end{tabularx}
\label{tab:topic_diversity_nodebunk}
\end{table}

\newpage
\begin{table}[H]
\centering
\caption{Estimation results for news lifetime [Column (1)] and post count [Column (2)]. The potential debunking posts are excluded during estimation. Reported are coefficient estimates with standard errors in parentheses. \sym{*} \(p<0.05\), \sym{**} \(p<0.01\), \sym{***} \(p<0.001\).}
\begin{tabularx}{\columnwidth}{@{\hspace{\tabcolsep}\extracolsep{\fill}}l*{2}{S}}
\toprule
&\multicolumn{1}{c}{(1)}&\multicolumn{1}{c}{(2)}\\
&\multicolumn{1}{c}{News lifetime}&\multicolumn{1}{c}{Post count}\\
\midrule
Falsehood   &       1.307\sym{***}&       2.234\sym{***}\\
            &     (0.049)         &     (0.087)         \\
Conspiracy  &       0.401\sym{*}  &       0.463         \\
            &     (0.203)         &     (0.344)         \\
Diversity   &      -0.004         &      -0.023         \\
            &     (0.005)         &     (0.026)         \\
Falsehood $\times$ Diversity&      -0.065         &       0.077         \\
            &     (0.044)         &     (0.093)         \\
Words     &      -0.078\sym{***}&      -0.619\sym{***}\\
            &     (0.010)         &     (0.029)         \\
Emotions       &       \checkmark&       \checkmark\\
Language    &       \checkmark  &       \checkmark  \\
MonthYear        &       \checkmark  &       \checkmark  \\
Intercept   &       2.599\sym{***}&       2.117\sym{***}\\
            &     (0.288)         &     (0.198)         \\
\midrule
\(N\)       &        \num{7097}         &       \num{7097}  \\
\(R^{2}\)   &       {0.496}         &           {0.191}         \\
\bottomrule
\end{tabularx}
\label{tab:news2tweets_nodebunk}
\end{table}

\begin{table}[H]
\centering
\caption{Estimation results for repost count [Column (1)], like count [Column (2)], and reply count [Column (3)]. The potential debunking posts are excluded during estimation. Reported are coefficient estimates with standard errors in parentheses. \sym{*} \(p<0.05\), \sym{**} \(p<0.01\), \sym{***} \(p<0.001\).}
\begin{tabularx}{\columnwidth}{@{\hspace{\tabcolsep}\extracolsep{\fill}}l*{3}{S}}
\toprule
&\multicolumn{1}{c}{(1)}&\multicolumn{1}{c}{(2)}&\multicolumn{1}{c}{(3)}\\
&\multicolumn{1}{c}{Repost count}&\multicolumn{1}{c}{Like count}&\multicolumn{1}{c}{Reply count}\\
\midrule
Falsehood   &      -0.207\sym{**} &      -0.050         &       0.525\sym{***}\\
            &     (0.065)         &     (0.065)         &     (0.060)         \\
Conspiracy  &       0.295\sym{***}&       0.320\sym{***}&       0.365\sym{***}\\
            &     (0.069)         &     (0.063)         &     (0.059)         \\
Diversity&       0.177\sym{***}&       0.191\sym{***}&       0.138\sym{***}\\
            &     (0.032)         &     (0.030)         &     (0.030)         \\
Falsehood $\times$ Diversity&      -0.118\sym{***}&      -0.086\sym{*}  &      -0.101\sym{**} \\
            &     (0.036)         &     (0.034)         &     (0.033)         \\
Words     &       0.527\sym{***}&       0.471\sym{***}&       0.454\sym{***}\\
            &     (0.017)         &     (0.016)         &     (0.015)         \\
Media       &       0.768\sym{***}&       0.581\sym{***}&       0.416\sym{***}\\
            &     (0.029)         &     (0.028)         &     (0.027)         \\
Verified    &       3.071\sym{***}&       3.127\sym{***}&       2.420\sym{***}\\
            &     (0.038)         &     (0.037)         &     (0.035)         \\
AccountAge  &       0.097\sym{***}&       0.083\sym{***}&       0.036\sym{**} \\
            &     (0.013)         &     (0.012)         &     (0.012)         \\
Followers   &       0.171\sym{***}&       0.174\sym{***}&       0.208\sym{***}\\
            &     (0.015)         &     (0.014)         &     (0.013)         \\
Followees   &       0.604\sym{***}&       0.440\sym{***}&       0.238\sym{***}\\
            &     (0.025)         &     (0.023)         &     (0.016)         \\
Emotions       &       \checkmark&       \checkmark&       \checkmark\\
Language       &       \checkmark&       \checkmark&       \checkmark\\
MonthYear       &       \checkmark&       \checkmark&       \checkmark\\
Intercept  &       0.160         &       1.050\sym{***}&      -1.452\sym{***}\\
            &     (0.174)         &     (0.159)         &     (0.159)         \\
\midrule
News-level REs   &       {\checkmark}         &          {\checkmark}           &     {\checkmark} \\
\midrule
\(N\)       &        \num{66588}         &       \num{66588}         &       \num{66588}         \\
\(R^{2}\)   &       {\xmark}         &          {\xmark}           &       {\xmark}         \\
\bottomrule
\end{tabularx}
\label{tab:post_engagement_nodebunk}
\end{table}

\newpage
\section{Robustness Check With BERTopic}
\label{sec:bertopic}

To ensure the generalization of our results across the topic modeling methods, we use BERTopic, a popular unsupervised topic modeling tool, as an alternative method to automatically cluster topics.\footnote{\url{https://maartengr.github.io/BERTopic/index.html}} The BERTopic model generates \num{1526} topic clusters for the whole news claims and posts. We use the topic clusters and their corresponding probabilities to calculate topic diversity again and repeat the analysis. Table \ref{tab:topic_diversity_bertopic} reports the estimation results for the topic diversity. Table \ref{tab:news2tweets_bertopic} reports the estimation results for the engagement at news source sharer level. Table \ref{tab:post_engagement_bertopic} reports the estimation results for the engagement at post viewer level. All the results are robust and consistently support our main findings.

\begin{table}[H]
\centering
\caption{Estimation results for the topic diversity in news claims [Column (1)], the topic diversity in news posts [Column (2)], and the changes in topic diversity from news claims to associated posts [Column (3)]. The topic diversity is calculated based on topics generated by BERTopic. Reported are coefficient estimates with standard errors in parentheses. \sym{*} \(p<0.05\), \sym{**} \(p<0.01\), \sym{***} \(p<0.001\).}
\begin{tabularx}{\columnwidth}{@{\hspace{\tabcolsep}\extracolsep{\fill}}l*{3}{S}}
\toprule
&\multicolumn{1}{c}{(1)}&\multicolumn{1}{c}{(2)}&\multicolumn{1}{c}{(3)}\\
&\multicolumn{1}{c}{News claim}&\multicolumn{1}{c}{News post}&\multicolumn{1}{c}{Claim $\rightarrow$ Post}\\
\midrule
Falsehood   &      -0.038         &      -0.042         &       0.059\sym{**} \\
            &     (0.042)         &     (0.028)         &     (0.021)         \\
Conspiracy  &      -0.150         &       0.046\sym{*}  &       0.045\sym{*}  \\
            &     (0.127)         &     (0.020)         &     (0.018)         \\
Words     &       0.089\sym{***}&       0.144\sym{***}&       0.110\sym{***}\\
            &     (0.013)         &     (0.005)         &     (0.004)         \\
Media       &                     &       0.016         &       0.017\sym{*}  \\
            &                     &     (0.009)         &     (0.007)         \\
Verified    &                     &       0.016         &      -0.016         \\
            &                     &     (0.011)         &     (0.009)         \\
AccountAge  &                     &       0.001         &       0.000         \\
            &                     &     (0.003)         &     (0.003)         \\
Followers   &                     &      -0.009\sym{*}  &      -0.007\sym{*}  \\
            &                     &     (0.004)         &     (0.004)         \\
Followees   &                     &      -0.019\sym{***}&      -0.016\sym{***}\\
            &                     &     (0.003)         &     (0.003)         \\
Emotions       &       \checkmark&       \checkmark&       \checkmark\\
Language       &       \checkmark&       \checkmark&       \checkmark\\
MonthYear       &       \checkmark&       \checkmark&       \checkmark\\
Intercept  &      -0.162         &       0.222\sym{***}&      -0.042         \\
            &     (0.160)         &     (0.051)         &     (0.044)         \\
\midrule
News-level REs&{\xmark}&{\checkmark}&{\checkmark}\\
\midrule
\(N\)       &        \num{7273}         &       \num{70904}         &       \num{70904}         \\
\(R^{2}\)   &       {0.041}         &          {\xmark}           &       {\xmark}              \\
\bottomrule
\end{tabularx}
\label{tab:topic_diversity_bertopic}
\end{table}

\newpage
\begin{table}[H]
\centering
\caption{Estimation results for news lifetime [Column (1)] and post count [Column (2)]. The topic diversity is calculated based on topics generated by BERTopic. Reported are coefficient estimates with standard errors in parentheses. \sym{*} \(p<0.05\), \sym{**} \(p<0.01\), \sym{***} \(p<0.001\).}
\begin{tabularx}{\columnwidth}{@{\hspace{\tabcolsep}\extracolsep{\fill}}l*{2}{S}}
\toprule
&\multicolumn{1}{c}{(1)}&\multicolumn{1}{c}{(2)}\\
&\multicolumn{1}{c}{News lifetime}&\multicolumn{1}{c}{Post count}\\
\midrule
Falsehood   &       1.209\sym{***}&       2.175\sym{***}\\
            &     (0.044)         &     (0.084)         \\
Conspiracy  &       0.405\sym{*}  &       0.553         \\
            &     (0.186)         &     (0.350)         \\
Diversity   &      -0.026\sym{***}&      -0.178\sym{***}\\
            &     (0.005)         &     (0.027)         \\
Falsehood $\times$ Diversity&      -0.302\sym{***}&      -0.117         \\
            &     (0.034)         &     (0.068)         \\
Words     &      -0.086\sym{***}&      -0.596\sym{***}\\
            &     (0.010)         &     (0.029)         \\
Emotions       &       \checkmark&       \checkmark\\
Language    &       \checkmark  &       \checkmark  \\
MonthYear        &       \checkmark  &       \checkmark  \\
Intercept   &       2.394\sym{***}&       2.029\sym{***}\\
            &     (0.275)         &     (0.202)         \\
\midrule
\(N\)       &        \num{7273}         &       \num{7273}  \\
\(R^{2}\)   &       {0.508}         &           {0.192}         \\
\bottomrule
\end{tabularx}
\label{tab:news2tweets_bertopic}
\end{table}

\begin{table}[H]
\centering
\caption{Estimation results for repost count [Column (1)], like count [Column (2)], and reply count [Column (3)]. The topic diversity is calculated based on topics generated by BERTopic. Reported are coefficient estimates with standard errors in parentheses. \sym{*} \(p<0.05\), \sym{**} \(p<0.01\), \sym{***} \(p<0.001\).}
\begin{tabularx}{\columnwidth}{@{\hspace{\tabcolsep}\extracolsep{\fill}}l*{3}{S}}
\toprule
&\multicolumn{1}{c}{(1)}&\multicolumn{1}{c}{(2)}&\multicolumn{1}{c}{(3)}\\
&\multicolumn{1}{c}{Repost count}&\multicolumn{1}{c}{Like count}&\multicolumn{1}{c}{Reply count}\\
\midrule
Falsehood   &      -0.134\sym{*}  &       0.000         &       0.521\sym{***}\\
            &     (0.060)         &     (0.060)         &     (0.056)         \\
Conspiracy  &       0.351\sym{***}&       0.387\sym{***}&       0.375\sym{***}\\
            &     (0.065)         &     (0.059)         &     (0.056)         \\
Diversity&       0.132\sym{***}&       0.197\sym{***}&       0.211\sym{***}\\
            &     (0.024)         &     (0.023)         &     (0.024)         \\
Falsehood $\times$ Diversity&      -0.058\sym{*}  &      -0.094\sym{***}&      -0.173\sym{***}\\
            &     (0.028)         &     (0.026)         &     (0.027)         \\
Words     &       0.537\sym{***}&       0.492\sym{***}&       0.450\sym{***}\\
            &     (0.016)         &     (0.015)         &     (0.014)         \\
Media       &       0.783\sym{***}&       0.591\sym{***}&       0.432\sym{***}\\
            &     (0.028)         &     (0.027)         &     (0.027)         \\
Verified    &       3.041\sym{***}&       3.076\sym{***}&       2.369\sym{***}\\
            &     (0.035)         &     (0.034)         &     (0.033)         \\
AccountAge  &       0.092\sym{***}&       0.086\sym{***}&       0.043\sym{***}\\
            &     (0.012)         &     (0.011)         &     (0.012)         \\
Followers   &       0.174\sym{***}&       0.181\sym{***}&       0.218\sym{***}\\
            &     (0.015)         &     (0.014)         &     (0.013)         \\
Followees   &       0.550\sym{***}&       0.389\sym{***}&       0.207\sym{***}\\
            &     (0.024)         &     (0.022)         &     (0.015)         \\
Emotions       &       \checkmark&       \checkmark&       \checkmark\\
Language       &       \checkmark&       \checkmark&       \checkmark\\
MonthYear       &       \checkmark&       \checkmark&       \checkmark\\
Intercept  &       0.013         &       0.919\sym{***}&      -1.506\sym{***}\\
            &     (0.168)         &     (0.154)         &     (0.152)         \\
\midrule
News-level REs&{\checkmark}&{\checkmark}&{\checkmark}\\
\midrule
\(N\)       &        \num{70904}         &       \num{70904}         &       \num{70904}         \\
\(R^{2}\)   &       {\xmark}         &          {\xmark}           &       {\xmark}              \\
\bottomrule
\end{tabularx}
\label{tab:post_engagement_bertopic}
\end{table}

\newpage
\section{Additional Analysis on Quote Count}
\label{sec:quote_count}

Table \ref{tab:post_quote_count} reports the full estimation results for quote count.

\begin{table}[H]
\centering
\caption{Estimation results for quote count. Reported are coefficient estimates with standard errors in parentheses. \sym{*} \(p<0.05\), \sym{**} \(p<0.01\), \sym{***} \(p<0.001\).}
\begin{tabularx}{.5\columnwidth}{@{\hspace{\tabcolsep}\extracolsep{\fill}}l*{1}{S}}
\toprule
&\multicolumn{1}{c}{(1)}\\
&\multicolumn{1}{c}{Quote count}\\
\midrule
Falsehood   &       0.102         \\
            &     (0.074)         \\
Conspiracy  &       0.585\sym{***}\\
            &     (0.084)         \\
Diversity&       0.186\sym{***}\\
            &     (0.039)         \\
Falsehood $\times$ Diversity&      -0.150\sym{***}\\
            &     (0.045)         \\
Words     &       0.414\sym{***}\\
            &     (0.021)         \\
Media       &       0.958\sym{***}\\
            &     (0.037)         \\
Verified    &       3.106\sym{***}\\
            &     (0.044)         \\
AccountAge  &       0.104\sym{***}\\
            &     (0.018)         \\
Followers   &       0.160\sym{***}\\
            &     (0.014)         \\
Followees   &       0.294\sym{***}\\
            &     (0.022)         \\
Emotions       &       \checkmark\\
Language       &       \checkmark\\
MonthYear       &       \checkmark\\
Intercept  &      -1.981\sym{***}\\
            &     (0.218)         \\
\midrule
News-level REs&{\checkmark}\\
\midrule
\(N\)       &        \num{70904}\\
\(R^{2}\)   &       {\xmark}\\
\bottomrule
\end{tabularx}
\label{tab:post_quote_count}
\end{table}

\end{document}